\documentclass[onecolumn]{els-mrw}          

\usepackage{amsmath,amssymb,amsfonts,amsthm,makeidx,graphicx}
\usepackage{bm,setspace,physics}
\usepackage{subfiles}
\usepackage[bookmarks=true,colorlinks=true,urlcolor=blue,linkcolor=blue,citecolor=blue,breaklinks]{hyperref}
\usepackage{txfonts}
\usepackage{helvet}

\begin{document}

\chapter{The Quantum Kicked Rotor: A Paradigm of Quantum Chaos \\
Foundational aspects and new perspectives}\label{chap1}

\author[1,2]{Giuliano Benenti}
\author[1,3]{Giulio Casati}%
\author[4,5,6]{Jiangbin Gong}%
\author[4,6]{Zhixing Zou}%

\address[1]{\orgname{Universit\`a degli Studi dell'Insubria}, \orgdiv{Center  for Nonlinear and Complex Systems, Dipartimento
di Scienza e Alta Tecnologia}, \orgaddress{via Valleggio 11, 22100 Como, Italy}}
\address[2]{\orgname{Istituto Nazionale di Fisica Nucleare}, \orgdiv{Sezione di Milano}, \orgaddress{via Celoria 16, 20133 Milano, Italy}}
\address[3]{\orgname{Federal University of Rio Grande do Norte}, \orgdiv{International Institute of Physics}, \orgaddress{Campus Universit\'ario - Lagoa Nova, CP. 1613,
Natal, Rio Grande Do Norte 59078-970, Brazil}}
\address[4]{\orgname{National University of Singapore}, \orgdiv{Department of Physics}, \orgaddress{Singapore}}
\address[5]{\orgname{MajuLab}, \orgdiv{CNRS-UCA-SU-NUS-NTU International Joint Research Unit}, \orgaddress{Singapore}}
\address[6]{\orgname{National University of Singapore}, \orgdiv{Centre for Quantum Technologies}, \orgaddress{Singapore}}


\maketitle

\begin{glossary}[Keywords]
Time scales of quantum chaos, correspondence principle, dynamical localization, quantum resonances, topological
aspects of kicked rotor variants, coupled kicked rotors, non-Hermitian kicked rotor
\end{glossary}

\begin{abstract}[Abstract]
The kicked rotor provides a simple yet powerful model for introducing many of the central concepts of classical and quantum chaos. Despite its apparent simplicity, it exhibits rich dynamical behavior and has found applications across a wide range of fields, including atomic and optical physics, condensed matter physics, and emerging quantum technologies.
This chapter begins by exploring foundational ideas using the kicked rotor as a unifying framework. We first discuss the transition from regular to chaotic motion in the classical system, and then introduce key quantum phenomena such as dynamical localization and quantum resonances. Special attention is devoted to the emergence of characteristic time scales and their role in the quantum–classical correspondence. To make these ideas more concrete, we also provide a brief overview of experimental realizations of the kicked rotor and its variants, illustrating how theoretical concepts are implemented in practice.
In the second part of the chapter, we guide the reader toward more recent and advanced developments. Topics include near-resonant dynamics, topological features of kicked systems, the emergence of quantum dynamical phases inferred from classical transport properties, and extensions to non-Hermitian physics. We conclude with a discussion of open problems and future perspectives, outlining directions in which the kicked rotor continues to offer valuable insights.
\end{abstract}


\section{Introduction}\label{sec:intri}

The general mathematical theory of dynamical systems—most notably ergodic theory~\cite{Cornfeld1982}—has its historical roots in classical mechanics. Its scope, however, is by no means confined to that setting and extends to a much broader class of phenomena, including systems governed by quantum mechanics. This immediately raises a fundamental question: how can classical dynamical chaos be understood within the framework of quantum mechanics?

In classical systems, chaos originates from strong local instabilities of motion, typically characterized by an exponential sensitivity to initial conditions. As a consequence, trajectories governed by deterministic equations become effectively unpredictable and exhibit behavior that appears random. The term \emph{chaotic} thus refers precisely to this loss of predictability: for almost all trajectories, reliable long-term prediction becomes impossible.
This notion can be formalized within the algorithmic theory of dynamical systems through the Alekseev--Brudno theorem (see Ref.~\cite{Alekseev1981}). According to this theorem, the information content $I(t)$ associated with a trajectory segment of length $t$—defined as the number of bits required to describe the orbit up to time $t$, or equivalently the length of the shortest program capable of reproducing those $t$ steps—grows asymptotically as
\begin{equation}
I(t) \sim h_{\mathrm{KS}}\, |t| ,
\end{equation}
where $h_{\mathrm{KS}}$ is the Kolmogorov--Sinai (KS) entropy, equal to the sum of all positive Lyapunov exponents~\cite{Ott2002}.

This result implies that predicting a new segment of a chaotic trajectory requires an amount of new information proportional to the length of that segment, independently of the previous history of the motion. In other words, no information about the future can be extracted from past observations: the future cannot be inferred from the past. This intrinsic unpredictability is the defining characteristic of chaotic dynamics.
By contrast, if the instability of motion grows only algebraically (i.e., sub-exponentially), the amount of new information required per unit time decreases with the length of the already known trajectory. In this case, the predictability improves with time, and asymptotically the dynamics becomes predictable.

It is important to emphasize, however, that short-term prediction is always possible, even in chaotic systems. This is true within a finite predictability window, $r < 1$, where
\begin{equation}
r = \frac{h_{\mathrm{KS}}\, |t|}{|\ln \mu|}.
\end{equation}
Here $\mu$ denotes the accuracy with which the trajectory is recorded. The regime $r \gg 1$ corresponds to effectively random behavior, where predictability is lost.

Exponential instability is typically associated with a continuous spectrum of motion, which in turn leads to the decay of correlations—a crucial prerequisite for the applicability of statistical descriptions. By contrast, systems with a discrete spectrum can exhibit at most linear instability in time~\cite{Casati1980}. Local exponential instability therefore defines a limiting regime known as \emph{dynamical chaos}. The opposite limiting case is that of regular (integrable) motion, characterized by a discrete spectrum.
This dichotomy poses a serious challenge for the understanding of quantum chaos. Indeed, for any quantum system whose motion is bounded in phase-space, both the energy and frequency spectra are necessarily discrete. According to the classical theory of dynamical systems, such behavior corresponds to the limiting case of regular motion and thus appears to be fundamentally incompatible with chaos.

The ultimate origin of this fundamental quantum property lies in the \emph{discreteness of phase-space itself}, which is a direct consequence of the Heisenberg uncertainty principle. This principle enforces a finite minimal volume for any elementary phase-space cell and underlies the entire structure of quantum mechanics. It is important to emphasize that the linearity of the Schr\"{o}dinger equation is not directly relevant in this context.
The requirement of nonlinearity for classical chaos pertains specifically to the trajectory-based description of motion, where nonlinearity is essential to confine otherwise linearly unstable trajectories. Indeed, the classical Liouville equation governing the evolution of phase-space distributions is itself linear. In mixing systems, any smooth distribution function relaxes, on average, toward a stationary state. This relaxation process is \emph{time-reversible}, just as the underlying microscopic dynamics is.
There is, however, a crucial distinction: while \emph{the phase-space distribution does not recur}, individual \emph{trajectories recur infinitely many times}—irrespective of whether the motion is regular or chaotic.

The time reversibility of the distribution function is associated with the increasingly complex structure it develops during relaxation. In the presence of exponential instability, the characteristic spatial scale of the oscillations in the distribution function decreases exponentially with time. The memory of the initial state is encoded in these progressively finer phase-space structures---\emph{a mechanism that relies fundamentally on the continuity of phase-space}.

To conclude, two properties are crucial for dynamical chaos in classical mechanics: \emph{a continuous spectrum of motion} and \emph{a continuous phase-space}. Both of these properties are absent in quantum mechanics. On the other hand, the correspondence principle requires transition from quantum to classical mechanics for all phenomena, including dynamical chaos. 

A resolution of this apparent contradiction (\emph{ad maiorem correspondentiae principii gloriam}\footnote{For the greater glory of the correspondence principle. In a letter from Heisenberg to Pauli,      September 30, 1924.}) is to focus on \emph{finite-time statistical properties}. The key observation is that the distinction between discrete and continuous spectra becomes unambiguous only in the infinite-time limit. 
This insight already emerged in the earliest numerical studies of quantum chaos~\cite{casati1979}, most notably through the introduction of a remarkably simple  model: the \emph{kicked rotor}.

The structure of this chapter reflects the progressive broadening of the kicked rotor’s conceptual and physical scope across several areas of modern physics. We begin by recalling its foundational role as a paradigmatic model of classical and quantum chaos, where it provides deep insight into the characteristic time scales governing quantum chaotic dynamics. These include the short Ehrenfest time, after which a quantum wave packet spreads significantly and no longer follows classical trajectories, and the longer localization time, beyond which quantum interference suppresses classical diffusion. We then emphasize its profound connection to Anderson localization, which established the model as a bridge between dynamical systems and disordered quantum transport. Building on this legacy, we survey a selection of advances that illustrate how the model has evolved into a versatile platform for exploring a wide spectrum of contemporary problems.

We begin with the single-particle Hermitian framework, including its semiclassical description, which continues to provide essential intuition linking phase-space structures to quantum interference phenomena. We then discuss representative single-particle generalizations, such as doubled-kicking protocols, different kicking potentials, and spinful extensions. These developments demonstrate how the model naturally connects to central themes in modern condensed-matter and Floquet physics, including topological phases, Anderson transitions, and quantum Hall physics. The chapter subsequently turns to coupled kicked rotors, which constitute a natural Hermitian many-body extension and offer a controlled setting for investigating the dynamics of entanglement and the emergence of complexity in interacting driven systems. Finally, we address non-Hermitian generalizations, an increasingly active research direction motivated by the growing interest in open quantum systems. Owing to the remarkable breadth and rapid expansion of the field, our treatment in each section is necessarily selective; rather than aiming at exhaustive coverage, we highlight key conceptual developments and emerging perspectives shaping current research.


\section{Foundational aspects}\label{sec:foundational}

\subsection{Classical kicked rotor}
\label{sec:krotclassical}

The kicked rotor belongs to the class of \emph{periodically driven dynamical
systems} and is governed by the Hamiltonian
\begin{equation}
  H(\theta,I;\tau) =
  \frac{I^2}{2} + V(\theta)
  \sum_{j=-\infty}^{+\infty} \delta(\tau-jT) \,,
  \label{krotham}
\end{equation}
where $(I,\theta)$ are conjugate action-angle variables, with $0\leq\theta<2\pi$. This Hamiltonian consists of two terms,
$H(\theta,I;\tau)=H_0(I)+U(\theta;\tau)$, where $H_0(I)=I^2\!/2$ corresponds to the kinetic energy of a free rotor—a particle moving on a circle parametrized by the coordinate 
$\theta$—while 
\begin{equation}
  U(\theta;\tau) =
  V(\theta) \sum_j \delta(\tau-jT) 
\end{equation}
represents the kicking potential that acts on the particle instantaneously at discrete time intervals $T$. 
Therefore, we say that the dynamics described by Hamiltonian
(\ref{krotham}) is \emph{kicked}.
The corresponding Hamiltonian equations of motion are given by 
\begin{equation}
  \left\{
    \begin{array}{l}
      \displaystyle
      \dot{I} =
      -\frac{\partial{H}}{\partial\theta} =
      -\frac{d V(\theta)}{d \theta}
      \sum_{j=-\infty}^{+\infty} \delta(\tau-jT) \,,
    \\[2ex]
      \displaystyle
      \dot\theta =
      \frac{\partial{H}}{\partial{I}} = I \,.
    \end{array}
  \right.
\end{equation}
These equations can be integrated straightforwardly, yielding a stroboscopic map that describes the evolution from time
$lT^-$ (just before the $l$-th kick) to time $(l+1)T^-$
(just before the $(l+1)$-th kick):
\begin{equation}
  \left\{
    \begin{array}{l}
      \displaystyle
      \bar{I} = I + F (\theta) \,,
    \\[2ex]
      \displaystyle
      \bar\theta= \theta + T\bar{I} \,,
    \end{array}
  \right.
  \label{kickedmap}
\end{equation}
where $F(\theta)=-dV(\theta)/d\theta$ is the force acting 
on the particle. Note that the transformation $(\theta,I)\to (\bar{\theta},\bar{I})$ is \emph{area-preserving}, since 
the associated Jacobian determinant is equal to one.

In the following, we focus on the kicked rotor, for which $V(\theta)=k\cos\theta$, so that the force becomes
$F(\theta)=-dV(\theta)/d\theta=k\sin\theta $. By rescaling the action (momentum) variable via $I\to{J=TI}$, 
the classical dynamics is seen to depend only on the single
parameter $K=kT$. Indeed, in terms of the rescaled variables $(J,\theta)$ 
Eq.~(\ref{kickedmap}) becomes
\begin{equation}
  \left\{
    \begin{array}{l}
      \displaystyle
      \bar{J} = J + K \sin\theta \,,
    \\[2ex]
      \displaystyle
      \bar{\theta} = \theta + \bar{J} \,,
    \end{array}
  \right.
  \label{krotmap2}
\end{equation}
which is known as the \emph{Chirikov standard map}~\cite{Chirikov1969,Chirikov1979}.
This map can be studied on the cylinder ($J\in(-\infty,+\infty)$), or on the torus $0\le J< 2{\pi}L$,
where $L$ is an integer (the integer condition ensures that no discontinuities are introduced in the second equation of Eq.~(\ref{krotmap2}) when 
$J$ is taken modulo $2{\pi}L$).
Due to its apparent simplicity and intrinsic richness, the standard map plays a key role in the study of both classical and quantum chaos. 
It can be interpreted as the Poincar\'e surface-of-section map for a conservative system with two degrees of freedom. 
What makes the standard map nearly universal is that it provides a local approximation (in momentum) for a broad class of dynamical systems. Notable examples include the microwave ionization of Rydberg atoms~\cite{Casati1984, Casati1986b, Casati1986c, Casati1988, Bayfield1989} (see Sec.~\ref{sec:hydrogen}) and two-dimensional billiards~\cite{Borgonovi1996, Casati1999, Casati2000, Jiaozi2021, Jiaozi2022},  to name just a few.

\begin{figure}
\centering
\includegraphics[width=.85\textwidth]{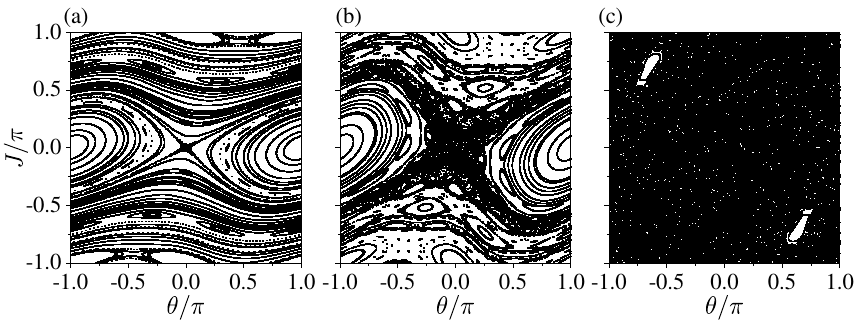}
\caption{Poincar\'e sections of the classical kicked-rotor map
\eqref{krotmap2} 
on the torus $\theta\in (-\pi,\pi]$, $J\in (-\pi,\pi]$,
for different kicking strengths: (a) $K=0.5$; (b) $K=0.97$; (c) $K=5$.
These plots are stroboscopic maps obtained by iterating the standard map for
different initial conditions.
For $K = 0.5$, the phase-space exhibits the main resonance at
$(\theta, J) = (\pi, 0)$, an unstable fixed point at $(0,0)$ asymptotic to the
separatrix, additional stable islands associated with higher-order resonances,
and deformed invariant tori that bound the motion in $J$.
As the system becomes increasingly nonlinear, the last surviving 
of these rotational invariant tori
are destroyed at $K \approx 0.97$.
Finally, for $K = 5$, the phase-space is predominantly chaotic, with only small
stability islands remaining.}
\label{fig:rotorspace}
\end{figure}

For  $K=0$, the motion is integrable, with the action $J$  being a constant of motion. The level curves, i.e., the lines  
$J(\theta)=J_0$, completely foliate the $(\theta,J)$ phase-space. These are called \emph{invariant tori}. 
On each torus, the motion is simply given by a rotation of 
$\theta$ with frequency $\omega=(\bar\theta-\theta)/{T}=J_0/T=I_0$.  
It is important to note that such rotations (or oscillations) are \emph{nonlinear}, since their frequency depends on the action. It is this nonlinearity which is responsible for chaos.
When the parameter $K$ is nonzero but sufficiently small, the Kolmogorov–Arnold–Moser (KAM) theorem predicts that 
some invariant tori—with strongly irrational frequencies—persist, albeit in a deformed state~\cite{Arnold1989,Lichtenberg2013} (see also \cite{Tomsovic2026} in this volume). These surviving tori constrain the motion in 
$J$, a situation referred to as \emph{global stability}.
For rational values of $\omega/(2\pi/T)=p/q$ (that is, $J_0=2\pi p/q$), the corresponding tori are destroyed. 
Of the infinite number of periodic trajectories that originally populated these tori, 
only 2$q$ periodic points survive. These are fixed points of the map iterated $q$ times. 
Half of these points are elliptic, while the other half are hyperbolic. 
This results in a \emph{mixed phase-space}, composed of chaotic regions and islands of stability
(see panel (a) in Fig.~\ref{fig:rotorspace}).
As $K$ increases, more and more tori are destroyed by the perturbation. At a critical value 
$K=K_c\approx 0.97$~\cite{Greene1979}, even the last invariant tori break down, and the motion in $J$ becomes unbounded
(see panel (b) in Fig.~\ref{fig:rotorspace}).
The fraction of phase-space occupied by stability islands decreases with increasing  $K$, and for $K\gtrsim 5$
it becomes negligible. The motion is then effectively fully chaotic and exhibits strong statistical properties, such as local instability, mixing, and rapid decay of correlations (see panel (c) in Fig.~\ref{fig:rotorspace}).

A less accurate but physically insightful estimate of $K_c$ can be obtained using
the Chirikov resonance-overlap criterion~\cite{Chirikov1969, Chirikov1979}. In the context of the kicked rotor,
we begin by applying the Poisson summation formula to the Dirac comb,
\begin{equation}
\sum_{j=-\infty}^{+\infty} \delta(\tau - jT)
= \frac{1}{T}\left(1 + 2\sum_{m=1}^{+\infty}
\cos\frac{2\pi m \tau}{T}\right),
\end{equation}
which allows the kicked-rotor Hamiltonian to be written as
\begin{equation}
H = \frac{J^2}{2}
+ K \sum_{m=-\infty}^{+\infty} \cos(\theta - 2\pi m t),
\label{eq:resonances}
\end{equation}
where $t \equiv \tau/T$ is a discrete time variable measured in units of map
iterations.
This Hamiltonian describes an infinite set of resonances located at
$J_m = 2\pi m$. The spacing between neighboring resonances is therefore
$\delta = 2\pi$. For each value of $m$, Eq.~(\ref{eq:resonances}) corresponds to an
effective pendulum, with a stable fixed point at
$(\theta, J) = (\pi, 2\pi m)$ and an unstable fixed point at
$(\theta, J) = (0, 2\pi m)$. The width of the associated separatrix is
$\Delta J = 4\sqrt{K}$.
According to the Chirikov criterion, the transition to global chaos occurs when
adjacent resonances overlap, i.e., when $\Delta J / \delta > 1$. This condition
yields
$K > \frac{\pi^2}{4} \approx 2.5$. 
Although this argument neglects higher-order resonances at
$J_{m\ell} = 2\pi m / \ell$ (with $\ell > 1$) and therefore overestimates the true critical value $K_c \approx 0.97$, it correctly captures the order of magnitude and provides an intuitive physical picture of the transition to chaos.


\emph{Local instability} can be conveniently studied by 
linearizing around a reference trajectory $\theta_0$ the standard map:
\begin{equation}
  \left[
    \begin{array}{c}
      \delta \bar{J} \\
      \delta \bar\theta
    \end{array}
  \right]
  = M
  \left[
    \begin{array}{c}
      \delta J \\
      \delta\theta
    \end{array}
  \right]
  =
  \left[
    \begin{array}{c@{\quad}c}
      1 &  K \cos\theta_0  \\
      1  & 1+ K \cos\theta_0
    \end{array}
  \right]
  \left[
    \begin{array}{c}
      \delta J \\
      \delta\theta
    \end{array}
  \right].
\end{equation}
The stability matrix $M$ has eigenvalues 
\begin{equation}
\mu_{\pm}(\theta_0)=1+\frac{K\cos\theta_0}{2}\pm\sqrt{K\cos\theta_0\left(1+\frac{K\cos\theta_0}{4}\right)}.
\end{equation}
Note that the eigenvalues depend not only on $K$ but also on $\theta_0$, and this implies local fluctuations in the instability along 
a trajectory. To compute the \emph{maximum Lyapunov exponent} $\lambda$ we need to average  $\ln|\mu_+|$ over the reference trajectory.
For $K\gg 1$, the trajectory essentially fills the entire phase-space and therefore $\lambda$ is obtained after phase averaging, i.e., by averaging 
$\ln|\mu_+(\theta_0)|\approx \ln(K|\cos\theta_0|)$ over $\theta_0$, yielding
\begin{equation}
\lambda\approx \ln(K/2).
\end{equation}

For any $K\gg 1$, the system exhibits \emph{normal diffusion} in the momentum variable. Although the standard map is a deterministic system, 
the motion of a trajectory along the momentum direction becomes, in practice, indistinguishable from a \emph{random walk}. As a result, the evolution of a distribution function
$f(J,t)$ is
governed by a \emph{Fokker--Planck
equation}:
\begin{equation}
  \frac{\partial{f}}{\partial{t}} =
  \frac{\partial}{\partial{J}}
  \left(\frac12 D \frac{\partial{f}}{\partial{J}} \right) .
  \label{fokkerplanck}
\end{equation}
Here 
$t$ is the discrete time and the diffusion coefficient $D$ is defined by
\begin{equation}
  D = \lim_{t\to\infty} \frac{\langle(\Delta{J}(t))^2\rangle}{t}, \,
\end{equation}
where $\Delta{J}\equiv{J}-\langle{J}\rangle$ and $\langle\dots\rangle$
denotes the average over an ensemble of trajectories. If at time
$t=0$ we take a phase-space distribution with initial momentum $J_0$ 
and random angles $0\leq\theta<2\pi$, then the solution 
of the Fokker--Planck equation (\ref{fokkerplanck}) is given by
\begin{equation}
  f(J,t) =
  \frac1{\sqrt{2 \pi D t}} \, \exp\left( -\frac{(J-J_0)^2}{2Dt} \right) .
\end{equation}
The width of this Gaussian distribution grows in time, according to
\begin{equation}
  \langle (\Delta{J}(t))^2 \rangle \approx D(K) \, t \,.
\end{equation}
For $K\gg 1$, the diffusion coefficient is well approximated by the random phase
approximation, in which we assume that there are no correlations between the
angles (phases) $\theta$ at different times. Hence, we have
\begin{equation}
  D(K) \approx
  \frac1{2\pi}\int_0^{2\pi} d\theta \, (\Delta{J}_1)^2 =
  \frac1{2\pi}\int_0^{2\pi} d\theta \, (K\sin\theta)^2 =
  \frac{K^2}{2},
\end{equation}
where $\Delta{J}_1=\bar{J}-J$ is the change in action after a single map step
\footnote{A more refined calculation~\cite{chirikov1989}, which takes into account phase correlations at different times, leads to
$D(K) \approx \frac{K^2}{2} C(K)$, where the correlator is given by 
$C(K)\approx 1-2 J_2(K)$, with $J_2$  denoting the Bessel function of the first kind of order 2. Note that $C(K)\to 1$ when $K\gg 1$, so that the random phase approximation
is recovered in this limit.}

\subsection{Quantum kicked rotor}

The quantized standard map~\cite{casati1979,Izrailev1990} 
(or quantum kicked rotor) 
is obtained by applying the usual canonical quantization rules: 
$\theta\to\hat{\theta}$ and $I\to\hat{I}=-i\partial/\partial\theta$, where we set $\hbar=1$.
The quantum evolution over one map iteration, that is, over one period of the perturbations, is described by a unitary operator 
$\hat{U}$, known as the \emph{Floquet operator}, which acts on the wave function 
$\psi$ as follows:
\begin{equation}
  \bar\psi =
  \hat{U}\,\psi = \hat{\cal{T}}
  \exp\bigg(
    -i \int_{lT^-}^{(l+1)T^-} {d\tau} \, \hat{H}(\theta,I;\tau)
  \bigg)\,
  \psi \,,
  \label{sawq}
\end{equation}
where $\hat{\cal{T}}$ is the time-ordering operator and 
$\hat{H}$ is the quantization of Hamiltonian (\ref{krotham}).
Since the potential $V(\theta)$ is
switched on only at discrete times $lT$, it is straightforward to obtain
\begin{equation}
  \bar\psi =
  e^{-i T \hat{I}^2\!/2} \, e^{-i{V}(\hat\theta)} \, \psi=
  e^{-i T \hat{I}^2\!/2} \, e^{-ik\cos\hat{\theta}} \, \psi =
  \hat{U}_T\hat{U}_k\,\psi.
  \label{krotquantum}
\end{equation}
The Floquet operator encapsulates all of the system’s dynamical properties and is expressed as the product of two non-commuting unitary operators, representing 
the effect of a single kick ($\hat{U}_k$) and 
the free rotation ($\hat{U}_T$), respectively \footnote{Although the operators $\hat{U}_k$ and $\hat{U}_T$ do not commute, 
exchanging them is simply equivalent to a shift in time.
A symmetric splitting of the Floquet operator, 
$\hat{U}=\hat{U}_{T/2}\hat{U}_k\hat{U}_{T/2}$ is sometimes used.}.
Note that, due to quantization, the dynamics depends essentially on both parameters
$k$ and $T$, and not just on their product $K$.
Since $[\hat{\theta},\hat{J}]=[\hat{\theta},T\hat{I}]=iT$, the effective Planck constant is given by
$\hbar_\mathrm{eff}=T$ 
and the classical limit corresponds to $T\to 0$  and $k\to\infty$
while keeping $K=kT$ constant.

Map (\ref{krotquantum}) can be made explicit in the coordinate (angular) representation as follows:
\begin{equation}
\hat{U}_k=\exp(- i k \cos\theta),\quad
\hat{U}_T=\exp\left(i \frac{T}{2} \frac{\partial^2}{\partial \theta^2}\right).
\label{eq:RkRT}
\end{equation}
In the momentum representation, defined by the Fourier transform
\begin{equation}
\psi(\theta) = \sum_{n = -\infty}^{+\infty} A_n \, e^{i n \theta},
\end{equation}
we obtain
\begin{equation}
\langle n|\hat{U}_k|m\rangle=\sum_{m = -\infty}^{+\infty} (-i)^{n - m} J_{n - m}(k),\quad
\langle n|\hat{U}_T|m\rangle= \exp\left(-i \frac{ T}{2} m^2 \right)\delta_{nm},
\label{eq:RkRT2}
\end{equation}
where $J_{n-m}$ is a Bessel function of the first kind. 
Since these functions rapidly decrease as the difference between the indices and the argument increases (for $|n-m|>k$), each kick couples approximately $k$
momentum eigenstates (which are also eigenstates of the unperturbed evolution at 
$k=0$).

\subsubsection{Quantum resonances}

Quantum evolution is governed by two distinct time scales: the first, $T$, corresponds to the period of the external perturbation; the second is $4\pi$, which arises from the property of the angular momentum operator whose spectrum, for $k=0$, consists of integers, i.e., $I_n=n$ (always at $\hbar=1$).
As a result, the Floquet operator remains unchanged under the transformation
$T\to T+4\pi$, see the expression for $\hat{U}_T$ in Eq.~(\ref{eq:RkRT2}).
Therefore, without loss of generality one can restrict 
to $0<T\le 4\pi$.
A resonance condition--without classical analogue--occurs when 
the ratio between the two characteristic time scales defined above is a 
rational number: 
\begin{equation}
    T=4\pi\,\frac{p}{q},
\end{equation}
with $p$ and $q$ integers.
On physical grounds, this condition is equivalent to stating that the energy carried by an integer number of photons of the perturbation, each with energy $\hbar \omega = 2\pi \hbar / T$, exactly matches the energy required for a transition between levels of the unperturbed rotor, whose eigenenergies are $E_n = n^2 \hbar^2 / 2$.
In this case, the rotor’s average energy grows quadratically in time.
The average energy after $t$ kicks is given by
\begin{equation}
E(t) = \frac{1}{2\pi} \int_0^{2\pi} d\theta\, \psi^*(\theta, tT) \left( -\frac{1}{2} \frac{\partial^2}{\partial \theta^2} \right) \psi(\theta, tT).
\end{equation}
Choosing $a\equiv p/q=1$, the wave function at time  $tT$ takes the form
\begin{equation}
\psi(\theta, tT) = \exp\left( -i k t \cos\theta \right) \psi(\theta, 0),
\end{equation}
from which it follows that
\begin{equation}
E(t) \sim \frac{k^2}{4} t^2, \qquad \text{as } t \to \infty.
\end{equation}
This quadratic energy growth can be demonstrated for any rational value of $a$~\cite{Izrailev1980,Guarneri2009}, with the exception of the case $a = 1/2$, for which the dynamics is strictly periodic:
$\hat{U}^2 = \hat{I}$.
On the other hand, numerical evidence~\cite{Izrailev1980,Casati1986} strongly suggests that the time scale over which resonant energy growth becomes observable increases extremely rapidly with $q$.
In the general case where 
$T/(4\pi)$ is irrational, the phenomenon of dynamical localization--to be discussed later--emerges.

\subsection{The quantum kicked rotor and the correspondence principle}

The kicked rotor serves as an ideal model for illustrating the characteristic time scales of quantum chaos--up to which phenomena such as exponential instability and diffusion can be recovered, in accordance with the correspondence principle.

\subsubsection{Ehrenfest time scale}

In order to understand the existence of different time scales in quantum motion, we compare the classical and quantum evolution starting from the same initial conditions. According to the \emph{Ehrenfest theorem}, a quantum wave packet follows a bundle of classical trajectories as long as the packet remains sufficiently narrow. During this interval, the quantum evolution is exponentially unstable and appears random, mirroring the behavior of the underlying classical dynamics.

However, the initial size of the quantum wave packet is bounded from below by the elementary quantum phase-space cell, which is of order $\hbar$. Let us consider an initial minimum-uncertainty wave packet of size $\Delta \theta_0 \Delta J_0 = T \Delta \theta_0 \Delta I_0 \sim \hbar_{\mathrm{eff}}$,
with
$\Delta \theta_0 \sim \Delta J_0 \sim \sqrt{\hbar_{\mathrm{eff}}}$,
which corresponds to the optimal, least-spreading wave packet. Due to classical exponential instability, the angular spread grows as 
$\Delta \theta(t) \sim \Delta \theta_0 \exp(\lambda t) \sim \sqrt{\hbar_{\mathrm{eff}}} \, e^{\lambda t}$,
where $\lambda$ is the maximum Lyapunov exponent of the system. Complete spreading over the angle variable \( \theta \) occurs after the so-called 
\emph{Ehrenfest time scale}:
\begin{equation}
t_E \sim \frac{1}{\lambda} \ln \left( \frac{1}{\hbar_{\mathrm{eff}}} \right)
\sim \frac{|\ln T|}{\ln(K/2)}.
\label{eq:tehrenfest}
\end{equation}
Thus, true dynamical chaos, characterized by exponential sensitivity to initial conditions, is limited in quantum mechanics (for Hamiltonian systems) to the logarithmically short Ehrenfest time scale. This phenomenon was first discussed in detail in Ref.~\cite{Berman1978}. Note that $t_E$ increases indefinitely as $\hbar_{\mathrm{eff}} \to 0 $, in agreement with the correspondence principle.

\subsubsection{Heisenberg time scale and dynamical localization}

The second characteristic time scale, $t^\star$--often referred to as the \emph{Heisenberg time}---marks the point at which the quantum evolution begins to diverge from classical diffusion. This departure is associated with the phenomenon of quantum dynamical localization, first observed in \cite{casati1979}.

\begin{figure}
\centering
\includegraphics[width=.40\textwidth]{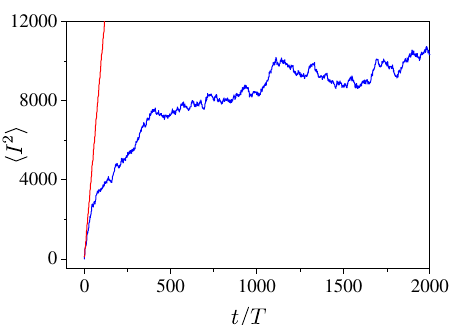}
\caption{Quantum localization (blue curve) of classical diffusion (red curve) in the kicked-rotor model. Both classical and quantum evolutions are computed using the parameter values $T = 0.25$ and $k = 20$, corresponding to an effective Planck constant of
$\hbar_\mathrm{eff} = 0.25$.
The classical dynamics is obtained by iterating the Chirikov standard map, starting at time $t = 0$ with an ensemble of $10^6$ initial state ssampled from a Gaussian distribution with standard deviation $\sqrt{\hbar_\mathrm{eff}/2}$ centered at $(0,0)$.
The quantum evolution is simulated by iterating the quantum kicked-rotor map, initialized with a minimum-uncertainty Gaussian wave packet of width $\Delta J = \Delta \theta = \sqrt{\hbar_\mathrm{eff}/2}$ , also centered at $(0,0)$.}
\label{fig:localization}
\end{figure}

\begin{figure}
\centering
\includegraphics[width=.40\textwidth]{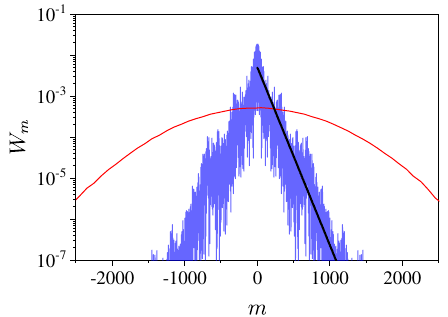}
\caption{Classical vs. quantum momentum distributions in the kicked rotor.
Classical (red curve) and quantum (blue curve) probability distributions over the momentum basis are shown for the kicked-rotor model at time $t = 3000 \gg t^\star \sim D_m \approx k^2/2 \sim 2 \times 10^2$. The dashed curve represents the classical Gaussian distribution, while the solid curve exhibits exponential localization in the quantum case.
The straight line corresponds to the theoretical prediction given by Eq.~(\ref{eq:locdistribution}), with localization length $\ell = D_m$. Parameter values and initial conditions are identical to those used in Fig.~\ref{fig:localization}.}
\label{fig:locdistr}
\end{figure}

For times $t > t^\star$, while the classical distribution continues to diffuse indefinitely, the quantum distribution instead reaches a stationary state. This steady state is exponentially localized in the momentum eigenbasis, as illustrated in Figs.~\ref{fig:localization} and \ref{fig:locdistr}:
\begin{equation}
W_m \equiv |\langle m | \psi \rangle|^2 \approx \frac{1}{\ell} \exp\left( -\frac{2|m - m_0|}{\ell} \right),
\label{eq:locdistribution}
\end{equation}
where $m$ labels the eigenstates of the momentum operator $\hat{I}$, such that 
$\hat{I} |m\rangle = m |m\rangle $, and $m_0 $ is the initial momentum. The parameter $\ell$, known as the \textit{localization length}, characterizes the width of the localized momentum distribution.

An estimate of the relaxation time $t^\star$ and the localization length $ \ell $ can be obtained through the following argument \cite{Chirikov1981}. 
During the initial stages of the evolution, the quantum dynamics closely follows classical diffusion. As a result, the number of unperturbed energy levels significantly involved in the dynamics increases with time according to
$\Delta m(t) \approx \sqrt{D_m t}$,
where 
$D_m\approx {k^2}/{2} $ 
is the classical diffusion coefficient, measured in terms of the number of momentum eigenstates.
Since the number of involved levels grows diffusively, as $\sqrt{t}$, and because the Heisenberg uncertainty principle limits energy resolution to $\Delta E \sim 1/t $, the discreteness of the spectrum eventually becomes dynamically relevant. 
This is the fundamental reason why localization occurs.

The localization length \( \ell \) can be estimated as follows. The localized quantum wave packet has significant amplitude over approximately $\ell$ basis states--both in the basis of momentum eigenstates and in the eigenbasis of the Floquet operator $\hat{U} $, which governs the time evolution via
$
|\bar{\psi}\rangle = \hat{U} |\psi\rangle.
$
Since $\hat{U}$ is unitary, its eigenvalues can be written in the form $\exp(i \epsilon_i) $, where the so-called \textit{quasi-energies} $\epsilon_i $ lie in the interval $[0, 2\pi)$. The mean level spacing between quasi-energy eigenstates that significantly contribute to the dynamics is thus approximately
$
\Delta E \approx {2\pi}/{\ell}.
$
According to the energy-time uncertainty principle, the minimum time required to resolve this spacing is given by the Heisenberg time:
\begin{equation}
t^\star \sim \frac{1}{\Delta E} \sim \ell.
\label{eq:Heisenbergtime}
\end{equation}
This equation defines the \textit{break time}, after which quantum features become dominant and the classical diffusive behavior is suppressed.
Up to time $t^\star$, diffusion involves a number of levels estimated as
\begin{equation}
\langle (\Delta m)^2 \rangle \approx D_m t^\star \approx \ell^2.
\label{eq:Deltam2}
\end{equation}
Combining Eqs.~(\ref{eq:Heisenbergtime}) and (\ref{eq:Deltam2}), 
we obtain the key relation
\begin{equation}
t^\star \approx \ell \approx D_m.
\label{eq:Siberia}
\end{equation}
Thus, the quantum localization length $\ell$ for the average probability distribution is approximately equal to the classical diffusion coefficient.
In accordance with the correspondence principle, this time scale diverges in the semiclassical limit:
\begin{equation}
t^\star \sim \frac{k^2}{2}\sim\frac{1}{\hbar_{\text{eff}}^2},
\end{equation}
where the final scaling follows from the relation $\hbar_{\text{eff}} = T \sim 1/k$ at fixed $K$.
Moreover, $t^\star \gg t_E$, where $t_E$ is the Ehrenfest time. This implies that classical-like diffusion can persist well beyond the Ehrenfest time, even in the absence of exponential instability.

\begin{figure}
\centering
\includegraphics[width=.40\textwidth]{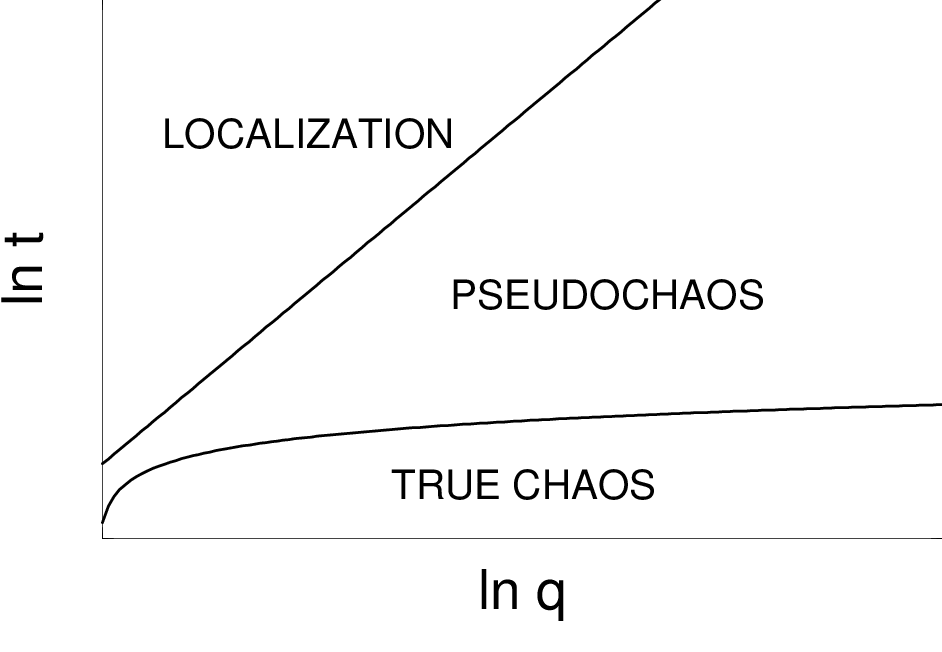}
\caption{Schematic representation of the characteristic time scales in the kicked-rotor model as a function of the quasi-classical parameter $q = 1/\hbar_{\text{eff}}$. The lower curve represents the Ehrenfest time scale $t_E \sim \frac{1}{\lambda} \ln q $, while the upper curve shows the Heisenberg (or localization) time scale $t^\star \sim q^2 $. In the semiclassical limit $\hbar_{\text{eff}} \to 0 $ (i.e., $q \to \infty$), both time scales diverge, in agreement with the correspondence principle.}
\label{fig:timescales}
\end{figure}

The concept of characteristic time scales in quantum dynamics reconciles the absence of dynamical chaos (as defined in ergodic theory) in quantum mechanics with the correspondence principle. The crucial point is that the following two limits do not commute:
\begin{equation}
\lim_{|t| \to \infty} \lim_{\hbar_{\text{eff}} \to 0} \neq \lim_{\hbar_{\text{eff}} \to 0} \lim_{|t| \to \infty}.
\end{equation}
While the left-hand side leads to classical chaos, the right-hand side results in an essentially quantum behavior with no chaos.
Both \textit{true chaos}-—characterized by exponential sensitivity to initial conditions—-and \textit{pseudochaos}---associated with quantum diffusion---are transient phenomena that manifest only over finite time intervals (see Fig.~\ref{fig:timescales}). The characteristic time scales over which these classical-like behaviors take place in the quantum dynamics of classically chaotic systems diverge in the semiclassical limit $\hbar_{\text{eff}} \to 0$, in agreement with the correspondence principle.

\subsection{Kicked rotor and Anderson localization}
\label{sec:Anderson}

Dynamical localization exhibits deep analogies with Anderson localization, which describes the suppression of electronic transport in disordered solids~\cite{Anderson1958}. 
In a random potential, a classical particle performs a random walk and eventually diffuses over a distance whose mean square grows linearly with time. It was first realized by Anderson that a quantum particle does not necessarily behave in this manner; depending on dimensionality and energy, it may instead remain localized indefinitely in the vicinity of its initial position.
In the context of Anderson localization, particular attention is given to so-called tight-binding models, which arise from the lattice discretization of the Schrödinger equation.
These models play a fundamental role in the study of transport properties in solids, particularly at low temperatures, where the electron wave function becomes highly sensitive to local impurities and imperfections in the crystal lattice.

In one dimension, the Anderson model is described by the eigenvalue equation
\begin{equation}
(H u)_n = W_n u_n + t (u_{n-1} + u_{n+1}) = E u_n,
\label{eq:anderson}
\end{equation}
where $u_n$ is the amplitude of the electronic wave function at lattice site $n$, $t$ denotes the hopping amplitude (i.e., kinetic energy term) between nearest-neighbor sites, and $W_n$ are random on-site energies. 
These random variables are assumed to be independently distributed. A common choice is to take $W_n$ uniformly distributed over the interval $[-W/2, W/2]$, where $W$ sets the disorder strength. A remarkable property of the one-dimensional Anderson model is that all eigenfunctions are exponentially localized, regardless of how small the disorder strength $W$ is (see, \textit{e.g.}, Ref.~\cite{Lee1985}). For a given eigenstate, the wave function decays as
\begin{equation}
u_n \propto \exp\left( -\frac{|n - n_0|}{\ell} \right),
\end{equation}
where $\ell$ is the localization length and $n_0$ is the center of localization (we assume the lattice size $N\gg \ell$).

Localization arises as a combined result of quantum tunneling and interference effects. A qualitative argument can be formulated as follows:
Consider a disordered potential as a sequence of potential steps with varying heights, separated by regions of zero potential. If the electron is initially located in the $n$-th region, it has a probability $p_t(n)$ of being transmitted to the $(n+1)$-th region, and a probability $p_r(n)$ of being reflected by the potential barrier, such that $p_t(n) + p_r(n) = 1$.
This process bears similarities to a classical random walk in configuration space, leading to diffusion. However, in quantum mechanics, interference effects of the wave packet play a crucial role: repeated coherent backscattering from multiple barriers leads to interference among all possible paths. 
As a consequence, the typical transmission probability through a large number of barriers $n \gg 1$ decays exponentially with $n$, indicating that the electron becomes localized in a region of size $\sim \ell$ around its initial position. This phenomenon---known as Anderson localization---arises from the interplay between quantum coherence and disorder.

In both the kicked rotor and the Anderson model, quantum interference effects inhibit unbounded diffusion—occurring in momentum space in the former, and in real space in the latter. This analogy can be formalized through a transformation that maps the quantum kicked rotor onto a disordered tight-binding model~\cite{Fishman1982,Grempel1994}.
Consider the eigenvalue equation for the Floquet operator of the kicked rotor:
\begin{equation}
\hat{U} |\psi\rangle = e^{-i \hat{H}_0 T} e^{-i \hat{V}} |\psi\rangle 
= e^{-i \varepsilon T} |\psi\rangle, 
\end{equation}
where $\hat{H}_0 = \hat{I}^2/{2}$ is the unperturbed Hamiltonian, and the kicking potential is given by $V(\theta) = k \cos \theta $. 
To proceed, we define two operators $\hat{W}$ and $\hat{t}$ through the following Cayley transformations:
\begin{equation}
e^{-i(\hat{H}_0 - \varepsilon) T} = \frac{1 + i \hat{W}}{1 - i \hat{W}}, 
\qquad 
e^{-i \hat{V}} = \frac{1 + i \hat{t}}{1 - i \hat{t}}, 
\end{equation}
which allows us to recast the Floquet eigenvalue problem into a more tractable algebraic form. Introducing the transformed state:
\begin{equation}
|\phi\rangle = (1 - i \hat{t})^{-1} |{\psi}\rangle,
\end{equation}
we find that the eigenvalue equation reduces to:
\begin{equation}
(\hat{W} + \hat{t}) |\phi\rangle = 0.
\end{equation}
We now expand the state $|\phi\rangle$ over the momentum eigenbasis 
$ \{ |{m}\rangle \}$:
\begin{equation}
|\phi\rangle = \sum_{m = -\infty}^{+\infty} u_m |m\rangle, \qquad 
\langle \theta | m \rangle = \frac{1}{\sqrt{2\pi}} e^{i m \theta}.
\end{equation}
Substituting this expansion into the transformed eigenvalue equation 
$(\hat{W} + \hat{t}) |\phi\rangle = 0 $, we obtain a tight-binding–like recursion relation:
\begin{equation}
W_n u_n + \sum_{\substack{l = -\infty \\ (l \ne 0)}}^{+\infty} t_l \, u_{n+l} = E u_n,
\label{eq:tightbinding}
\end{equation}
where the site energies $W_n$ and hopping amplitudes $t_l$ are given by:
\begin{align}
W_n = \tan \left( \frac{1}{2} \left( \varepsilon T - \frac{1}{2} n^2 T \right) \right),
\quad
t_l 
= -\frac{1}{2\pi} \int_0^{2\pi} d\theta \, e^{i l \theta} \tan \left( \frac{V(\theta)}{2} \right),
\end{align}
with  $E_ \equiv -t_0$.
The tight-binding model (\ref{eq:tightbinding}) describes the motion of an electron in a lattice whose sites are indexed by $n$, site energies $ W_n$ and hopping amplitudes $t_l$.
In contrast to the standard Anderson model (\ref{eq:anderson}), the hopping terms here are not restricted to nearest-neighbor sites.

It is important to note that, unlike in the Anderson model, the site energies $W_n $ are not drawn from a random distribution but are instead determined by the system's dynamics. Nevertheless, if the ratio $T / (4\pi) $ is irrational, the sequence $\{ W_n \} $ becomes effectively pseudo-random, leading to Anderson-like localization. This correspondence has been confirmed by several numerical studies showing exponential localization of the eigenfunctions when $ T $ is an irrational multiple of $4\pi$. On the other hand, when $ T / (4\pi)$ is a rational number, the site energies $ W_n $ become periodic. In this case, the eigenfunctions are extended Bloch waves, and the particle propagates freely through the lattice. In the context of the kicked rotor, this corresponds to the quantum resonances discussed earlier.

It should be emphasized that, in the dynamical case:  
\begin{itemize}
    \item[(i)] Localization occurs in \emph{momentum space} rather than in configuration space, and is associated with the \emph{quasi-energy eigenfunctions} of the Floquet operator;
    \item[(ii)] No external randomness is introduced; the kicking potential is strictly \emph{periodic} in time.
\end{itemize}
For these reasons, the localization phenomenon arising from quantum interference in classically chaotic systems is referred to as \emph{dynamical localization}, to distinguish it from \emph{Anderson localization}, which originates from static disorder in the Hamiltonian \footnote{We do not address here the question of how random the sequence $W_n$ must be for Anderson localization to occur. This issue is related to the number-theoretic properties of $T$.}.

\subsection{A brief sketch of the field-theoretical approach}

A unified field-theoretical framework has been developed to describe dynamical localization and quantum resonances in the kicked rotor~\cite{Altland1993,Altland1996,Tian2010,Altland2015}. 
This approach represented an important advance, replacing phenomenological understanding with a microscopic theory whose domain of validity is explicitly specified. In doing so, it reveals a precise correspondence between chaotic systems exhibiting dynamical localization and disordered metals.
The starting point is the Floquet operator introduced in Eq.~\eqref{sawq}. To investigate the long-time dynamics, we consider the transition probability $P_t(n,n')=|\langle n|\hat U^t|n'\rangle|^2$ between angular-momentum states $|n'\rangle$ and $|n\rangle$ after $t$ kicks. 
Its frequency representation defines the density correlation function
\begin{equation}
K_\omega(n,n')=\sum_{t=0}^\infty e^{i\omega t}P_t(n,n'),
\label{eq:DCF}
\end{equation}
where $\omega$ carries an infinitesimal positive imaginary part ensuring convergence. This correlation function contains information about momentum spreading and the corresponding kinetic-energy growth.

To formulate $K_\omega$ in a field-theoretical framework, one introduces the retarded and advanced Floquet Green functions,
\begin{equation}
G^\pm(\omega_\pm)=\left[1-(e^{i\omega_\pm}\hat U)^{\pm1}\right]^{-1},
\label{eq:KR_green}
\end{equation}
where $\omega_\pm=\omega_0\pm\omega/2$. 
The retarded function $G^+$ describes forward Floquet propagation and the advanced function $G^-$ describes backward propagation.
The density correlation function in Eq.~\eqref{eq:DCF} can then be expressed as
\begin{equation}
K_\omega(n,n')=\left\langle\langle n|G^+(\omega_+)|n'\rangle\langle n'|G^-(\omega_-)|n\rangle\right\rangle_{\omega_0},
\label{eq:KR_green_corr}
\end{equation}
where $\langle\cdots\rangle_{\omega_0}$ denotes an average over the reference quasienergy $\omega_0\in[0,2\pi)$. 
This average selects contributions with equal forward and backward propagation times, without introducing disorder averaging.

The field-theoretical construction proceeds by representing the Green functions in Eq.~\eqref{eq:KR_green} as Gaussian integrals over auxiliary supervector fields containing both bosonic and fermionic components. The bosonic fields generate the Green functions, while the fermionic fields ensure the normalization of the functional integral. The resulting fields carry retarded--advanced, boson--fermion, and angular-momentum indices, s well as a time-reversal grading. 
The quasienergy average couples the retarded and advanced sectors; this coupling is handled exactly by the color-flavor transformation.
This transformation introduces two auxiliary supermatrix fields $Z$ and $\widetilde Z$, that encode correlations between forward and backward propagation. 
After integrating out the original supervector fields and retaining only the low-energy sector consistent with time-reversal symmetry, the action takes the form~\cite{Tian2010}
\begin{equation}
S[Z,\widetilde Z]=-\frac12\operatorname{Str}\ln(1-Z\widetilde Z)+\frac12\operatorname{Str}\ln(1-e^{i\omega}\hat U^\dagger Z\hat U\widetilde Z),
\label{eq:KR_Z_action}
\end{equation}
where $\operatorname{Str}$ denotes the supertrace. The fields satisfy the symmetry constraint $\widetilde Z=\tau^{-1}Z^T\tau$, with $\tau$ implementing the time-reversal structure. 
Importantly, the microscopic Floquet operator enters the action explicitly through $\hat{U}^\dagger Z\hat{U}$, preserving the connection to the underlying dynamics.

The long-time dynamics is controlled by configurations of small action, corresponding to fluctuations that change only weakly under Floquet evolution. Expanding Eq.~\eqref{eq:KR_Z_action} to quadratic order at $\omega=0$ gives
\begin{equation}
S^{(2)}[Z,\widetilde Z]=\frac12\operatorname{Str}\left[\widetilde Z(1-\mathcal P)Z\right],\qquad \mathcal P Z=\hat U^\dagger Z\hat U.
\label{eq:KR_quadratic}
\end{equation}
Here $\mathcal P$ is the Floquet evolution superoperator, which describes, in the Heisenberg picture, the evolution of an operator over one driving period. In the angular-momentum basis, its matrix elements are
\begin{equation}
\mathcal P_{nn';mm'}=U_{mn}^*U_{m'n'},\qquad (\mathcal P Z)_{nn'}=\sum_{m,m'}\mathcal P_{nn';mm'}Z_{mm'}.
\label{eq:KR_P_matrix}
\end{equation}
The soft modes correspond to configurations satisfying $\mathcal P Z\simeq Z$, which evolve only slowly over successive periods. In the strongly chaotic regime, these modes are approximately diagonal in angular-momentum space and vary slowly with $n$, namely $Z_{nn'}\simeq B(n)\delta_{nn'}$. The remaining rapidly varying, or massive, fluctuations can be integrated out.

The spatial gradient term of the effective theory follows directly from the microscopic Floquet dynamics. For the kicking operator $\exp(-ik\cos\hat\theta)$, the angular-momentum transition amplitudes are proportional to Bessel functions $J_r(k)$. Projecting Eq.~\eqref{eq:KR_P_matrix} onto the diagonal soft modes gives, to leading order,
\begin{equation}
(\mathcal PB)(n)\simeq\sum_r J_r^2(k)B(n+r).
\label{eq:KR_projected}
\end{equation}
For slowly varying fields, the long-wavelength expansion of Eq.~\eqref{eq:KR_projected} yields
\begin{equation}
1-\mathcal P(\varphi)\simeq D_0\varphi^2,\qquad D_0=\frac12\sum_r r^2J_r^2(k)=\frac{k^2}{4},
\label{eq:KR_diffusion}
\end{equation}
where $\varphi$ is conjugate to the angular-momentum coordinate and $D_0$ is the bare diffusion coefficient. Thus, the diffusive gradient term arises from the microscopic angular-momentum transitions. Integrating out the massive modes further renormalizes $D_0$ to the quantum diffusion coefficient $D_{\rm q}$.

To incorporate the nonlinear fluctuations of the soft modes, one introduces the supermatrix field
\begin{equation}
Q=g\Lambda g^{-1},\qquad g=\begin{pmatrix}1&B\\ \widetilde B&1\end{pmatrix},\qquad \Lambda = \begin{pmatrix}1&0\\0&-1\end{pmatrix}_{\mathrm{AR}},
\label{eq:KR_Q}
\end{equation}
where $\Lambda$ distinguishes the retarded and advanced sectors and $Q^2=1$. Expanding $Q$ around $\Lambda$ reproduces the quadratic diffusive action, while finite rotations capture the nonlinear fluctuations responsible for quantum interference. Combining the spatial gradient term with the leading frequency dependence gives the nonlinear sigma-model action~\cite{Tian2010}
\begin{equation}
S[Q]=\frac{1}{16}\int_0^q dn\,\operatorname{Str}\left[D_q(\mathcal D_nQ)^2-2i\omega Q\Lambda\right],
\label{eq:KR_sigma}
\end{equation}
where $\mathcal D_nQ=i\partial_nQ-[\hat\vartheta,Q]$ is the covariant derivative associated with the Bloch phases. 
This action has the same long-wavelength structure as the nonlinear sigma model describing Anderson localization in quasi-one-dimensional disordered conductors~\cite{Altland1996,Tian2010}. The derivation is controlled in the strongly chaotic, long-wavelength regime, with the relevant length scales exceeding the microscopic transport scale $\ell\sim k$.

For nonresonant values of the effective Planck constant, 
the momentum-space periodicity plays no role,
and Eq.~\eqref{eq:KR_sigma} reduces to the field theory of an infinitely extended disordered wire. At the Gaussian level, the density correlation function takes the diffusive form
\begin{equation}
K_\omega(\varphi)\simeq\frac{1}{-i\omega+D_q\varphi^2}.
\label{eq:KR_diffusive_propagator}
\end{equation}
This reproduces diffusive kinetic-energy growth, $E(t)\simeq D_{\rm q}t$, at intermediate times. At longer times, nonlinear quantum fluctuations suppress diffusion, producing $D(\omega)\propto-i\omega\xi^2$, where $\xi\sim D_{\rm q}$ is the localization length. Consequently, the kinetic energy saturates at $E(t\rightarrow\infty)\sim\xi^2$, corresponding to dynamical localization.

For rational values $\hbar_{\mathrm{eff}}/(4\pi)=p/q$, the Floquet operator is invariant under translations by $q$ angular-momentum states. The corresponding Bloch decomposition maps Eq.~\eqref{eq:KR_sigma} onto a disordered metallic ring of 
$q$ sites, with the Bloch phase playing the role of a fictitious Aharonov--Bohm flux. The interplay 
between $q$ and $\xi$ then determines the long-time behavior. When $\ell\ll q\ll\xi$, the dynamics becomes sensitive to the momentum-space periodicity before localization develops. At sufficiently long times, the spatially uniform sector of the sigma model dominates, leading to quantum-resonant ballistic energy growth,
\begin{equation}
E(t)\sim\frac{D_q}{q}t^2.
\label{eq:KR_resonant_energy}
\end{equation}
In contrast, when $q\gg\xi$, localization sets in before the periodicity is resolved, and resonant corrections are exponentially suppressed in $q/\xi$. Even in the localized regime, however, the residual narrow Bloch bands can produce ballistic spreading at extremely long times.
For exactly rational values, the residual narrow Bloch bands can nevertheless produce ballistic spreading at extremely long times.

\subsection{Experimental realizations: a brief overview}

\subsubsection{The hydrogen atom in a microwave field}
\label{sec:hydrogen}

One of the most significant cases where classical and quantum chaos were directly confronted is the explanation of the experiments on the microwave ionization of highly excited hydrogen atoms~\cite{Bayfield1974,Koch1995}. In these experiments, hydrogen atoms were prepared in very elongated Rydberg states with a high principal quantum number $n_0 \approx 63{-}69 $ and injected into a microwave cavity. The ionization rate was then measured as a function of the microwave field intensity.
The applied microwave frequency was $\omega = 9.9\,\mathrm{GHz}$, corresponding to a photon energy much smaller than the ionization energy of level $n = 66$, and even lower than the energy required for a single-photon transition between adjacent levels (e.g., from $n = 66$  to $n = 67$). Surprisingly, efficient ionization was observed when the electric field amplitude exceeded a threshold value $\varepsilon \approx 20\,\mathrm{V/cm}$, which is significantly lower than the static field required for Stark ionization.
The threshold behavior was explained within the framework of classical dynamics, as the critical field strength corresponding to the onset of chaotic diffusion in the classical action variable $ n$~\cite{Leopold1978,Jensen1982, Delone1983}.
However, the hydrogen atom remains a fundamentally quantum object, and it is possible to identify parameter regimes—analogous to those of the quantum kicked rotor—where classical chaotic diffusion is suppressed by quantum interference. This leads to the phenomenon of quantum localization, where ionization is effectively inhibited despite the presence of classically chaotic dynamics.

The Hamiltonian for a hydrogen atom interacting with a time-periodic, linearly polarized electric field, in the dipole approximation, is given by:
\begin{equation}
H = \frac{p^2}{2} - \frac{1}{r} + \varepsilon z \cos(\omega t),
\label{eq:hydrogen_hamiltonian}
\end{equation}
where \( \varepsilon \) and \( \omega \) are the amplitude and the frequency of the electric field, respectively, which is assumed to be directed along the \( z \)-axis \footnote{Note that we use atomic units, where fundamental constants are set to unity: the electron mass  $m = 1$, the elementary charge $e = 1$, and the reduced Planck constant $\hbar = 1$. 
}. When the electron occupies a state that is highly extended along the direction of the electric field, the one-dimensional model becomes a good approximation of the full three-dimensional motion. This model is described by the Hamiltonian:
\begin{equation}
H_{\text{1D}} = \frac{p^2}{2} - \frac{1}{z} + \varepsilon z \cos(\omega t), \quad z > 0.
\label{eq:H1D}
\end{equation}
The unperturbed Hamiltonian
\begin{equation}
H^{(0)}_{\text{1D}} = \frac{p^2}{2} - \frac{1}{z}
\end{equation}
describes both bound motion (for energies $E < 0$) and unbound motion 
(for $E > 0$). Since we are primarily interested in the dynamics preceding ionization, we restrict ourselves to negative energies and introduce action-angle variables $(n, \theta)$, leading to the Hamiltonian:
\begin{equation}
H_{\text{1D}} = -\frac{1}{2n^2} + \varepsilon z(n, \theta) \cos(\omega t).
\label{eq:H1D_action_angle}
\end{equation}

It is important to note that the classical dynamics depends only on the rescaled parameters
$\varepsilon_0 \equiv \varepsilon n_0^4$ and $ \omega_0 \equiv \omega n_0^3$,
where $n_0$ is the initial value of the action variable (corresponding, in the quantum case, to the principal quantum number of the initial state). To demonstrate this scaling behavior, we perform the following variable transformations:
\begin{equation}
z \rightarrow \frac{z}{n_0^2}, \quad t \rightarrow \frac{t}{n_0^3}, \quad
\varepsilon \rightarrow n_0^4 \varepsilon, \quad \omega \rightarrow n_0^3 \omega.
\label{eq:scaling_transformation}
\end{equation}
Under these transformations, the momentum scales as
$
p = \frac{dz}{dt} \rightarrow n_0 p,
$
and the Hamiltonian transforms as
$
H_{\text{1D}} \rightarrow n_0^2 H_{\text{1D}}.
$
Since the Hamiltonian is only multiplied by a constant factor, the equations of motion remain unchanged; only the size of the trajectories is rescaled.
If the initial unperturbed energy is
$
E = -\frac{1}{2n_0^2},
$
then the transformation (\ref{eq:scaling_transformation}) maps this to the energy of the first Bohr level:
$
E \rightarrow n_0^2 E = -\frac{1}{2}.
$
We also observe that the parameters $\varepsilon_0$  and $\omega_0$ represent, respectively, the ratio of the field strength $\varepsilon$ to the characteristic Coulomb field strength $\varepsilon_C = z^{-2} \sim n_0^{-4}$, and
the ratio of the driving frequency $ \omega$  to the unperturbed Kepler orbital frequency $\omega_K = n_0^{-3}$.
Finally, the rescaled time corresponds to the number of unperturbed Kepler periods (up to a factor of $ 2\pi$:
$
t/n_0^3 = 2\pi {t}/{T_K}$, with $T_K = 2\pi n_0^3$).

Scaling (\ref{eq:scaling_transformation}) is not valid in quantum mechanics.
Indeed, the canonical commutation relation is modified under such 
scaling:
$
[z, p] \rightarrow \left[ n_0^{-2} z,\, n_0 p \right] = i n_0^{-1}.
$
Therefore, in quantum mechanics an additional parameter appears, which is absent in classical mechanics: the effective Planck constant,
$
\hbar_{\text{eff}} = \frac{1}{n_0}.
$
The classical limit corresponds to $\hbar_{\text{eff}} \to 0 $, 
i.e., $n_0 \to \infty$, while keeping the rescaled parameters $\varepsilon_0$ and $\omega_0$  constant.

If the field frequency exceeds the electron frequency, $\omega_0 > 1$, it can be shown~\cite{Casati1988} that the dynamics of system~(\ref{eq:H1D_action_angle}) can be approximately described over one Kepler period by a discrete-time map, the so-called \emph{Kepler map}:
\begin{equation}
\begin{cases}
\bar{N} = N + k \sin \phi, \\
\bar{\phi} = \phi + \dfrac{\pi}{\sqrt{2 \omega}} (-\bar{N})^{-3/2},
\end{cases}
\label{eq:kepler_map}
\end{equation}
where $N = E / \omega$, $\phi$ is the field phase at perihelion, and the perturbation parameter is given by
\begin{equation}
k \approx \dfrac{2.6\, \varepsilon}{\omega^{5/3}}.
\label{eq:k_kepler_map}
\end{equation}
We obtain a map description because the external field exerts its dominant perturbing effect on the electron’s free Keplerian motion primarily when the electron passes near the perihelion. This behavior originates from the Coulomb singularity and results in a kick-like action of the external perturbation.
Note that the Kepler map is only defined for states with $N < 0$, i.e., bound states. If an iteration of the map leads to  $\bar{N} > 0$, 
then $E > 0$ and the electron is considered ionized.
By linearizing the second equation of~(\ref{eq:kepler_map}) around the initial value $ N_0 = -\dfrac{1}{2 n_0^2 \omega}$, the Kepler map reduces to the standard map:
\begin{equation}
\begin{cases}
\bar{N}_\phi = N_\phi + k \sin \phi, \\
\bar{\phi} = \phi + T N_\phi,
\end{cases}
\label{eq:standard_map}
\end{equation}
where $N_\phi = N - N_0$ (interpreted in the quantum case as the number of photons absorbed) and 
$T = 6\pi \omega^2 n_0^5$ \footnote{An irrelevant constant phase shift has been omitted in the second equation.}.

As discussed in Sec.~\ref{sec:krotclassical}, the transition to classical chaos occurs when $K = kT \approx 1 $, leading to the \emph{classical chaos border}:
\begin{equation}
\varepsilon_c \approx \dfrac{1}{49\, n_0^5\, \omega^{1/3}},
\label{eq:eps_classical_chaos}
\end{equation}
or in terms of rescaled units:
\begin{equation}
\varepsilon_{0c} \approx \dfrac{1}{49\, \omega_0^{1/3}}.
\label{eq:eps0_classical_chaos}
\end{equation}

On the quantum side, drawing from the kicked-rotor analogy, quantum localization is expected over a length (measured in number of photons):
\begin{equation}
\ell_\phi \approx \dfrac{k^2}{2} \approx \dfrac{3.3\, \varepsilon^2}{\omega^{10/3}}.
\label{eq:localization_length}
\end{equation}
The ionization process is thus governed by two characteristic scales:
\begin{itemize}
    \item[(i)] the localization length $\ell_\phi$, and
    \item[(ii)] the "sample size" $N_I$, i.e., the number of photons required to reach the continuum from the initial state $n_0$,
\end{itemize}
\begin{equation}
N_I = \dfrac{1}{2 n_0^2 \omega}.
\label{eq:sample_size}
\end{equation}
If $\ell_\phi \ll N_I $, quantum localization prevents ionization. Conversely, if $\ell_\phi > N_I$, chaotic diffusion proceeds until ionization.
The \emph{quantum delocalization border} is obtained by setting $\ell_\phi = N_I$, which yields:
\begin{equation}
\varepsilon_q \approx 0.4\, \dfrac{\omega^{7/6}}{n_0}.
\label{eq:quantum_border}
\end{equation}
In order to observe chaotic ionization, the condition $\varepsilon > \varepsilon_q$ must be satisfied.

\begin{figure}
\centering
\includegraphics[width=.40\textwidth,angle=270]{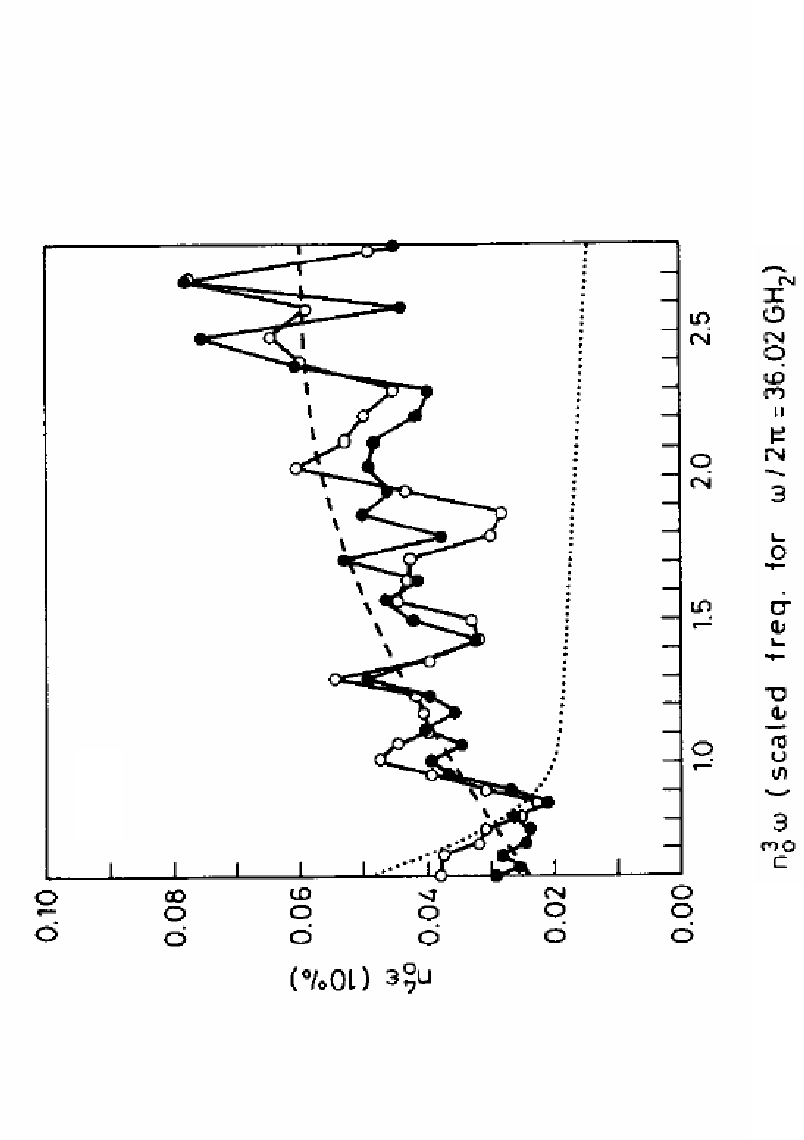}
\caption{Scaled 10\% ionization threshold field as a function of the rescaled frequency, from experimental data (empty circles, taken from~\cite{Bayfield1989}) and from numerical integration of the quantum Kepler map (filled circles). The dashed curve shows the quantum theoretical prediction based on localization theory, while the dotted curve represents the classical chaos border. Figure adapted from \cite{Casati1990}. 
}
\label{fig:hydrogen}
\end{figure}

A comparison between localization theory and experimental data~\cite{Bayfield1989} is shown in Fig.~\ref{fig:hydrogen}. The empty circles represent the experimentally observed threshold values of the microwave peak amplitude required to ionize 10\% of the atoms \footnote{This threshold corresponds to a fixed ionization probability and serves as a meaningful indicator for comparing theory and experiment.}. The dotted curve shows the classical chaos border, while the dashed curve represents the prediction of the localization theory for 10\% ionization. 
The filled circles indicate numerical results obtained from the integration of the quantized Kepler map. 
The agreement between experimental data and the numerical results is quite remarkable, especially considering that the Kepler map provides only an approximate description of the true quantum dynamics. Furthermore, the localization theory successfully captures the average behavior of the ionization threshold.

As shown in Fig.~\ref{fig:hydrogen}, both the numerical and experimental data exhibit noticeable deviations from the average theoretical prediction. These fluctuations are analogous to the conductance fluctuations observed in finite solid-state samples~\cite{Casati1990}. In the case of microwave ionization of hydrogen atoms, the system effectively forms a finite ``photonic lattice," since only a finite number of photons is required to ionize the atom. 
However, while in solid-state systems---typically modeled by the Anderson Hamiltonian---such fluctuations arise from the static randomness of the potential landscape, in the hydrogen atom they originate from the underlying dynamical chaos of the classical system. 

\subsubsection{Cold atoms in optical lattices}
\label{sec:coldatoms}

In this experimental implementation of the quantum kicked rotor model~\cite{Moore1995,Klappauf1999}, neutral sodium atoms were trapped and cooled using a magneto-optical trap. Then, the atoms were released from the trap and allowed to interact with a flashed optical lattice.

The dynamics is modeled in terms of two-level atoms with transition frequency 
$\omega_0$, interacting with a pulsed standing wave of linearly polarized, near-resonant light of frequency $\omega_L$. For sufficiently large detuning $\Delta = \omega_0 - \omega_L$ (relative to the natural linewidth), the excited-state amplitude can be adiabatically eliminated~\cite{Graham1992}. 
The atom can then be treated as a point particle, whose center-of-mass motion is described by a kicked-rotor-like Hamiltonian: 
\begin{equation}
H = \frac{P^2}{2M} + V_0 \cos(2k_L x) \sum_{n\in\mathbb{Z}} F(\tau - nT),
\end{equation}
where $V_0 = {\hbar \Omega^2}/{8\Delta}$, $k_L$ is the wave number of the light, $\Omega = {2\mu E_0}/{\hbar}$ is the resonant Rabi frequency, $\mu$ is the atomic dipole moment, $E_0$ is the electric field amplitude,
$F(\tau)$ represents a pulse centered at $\tau = 0$ with duration $\tau_p$, and $T$ is the period of the standing wave pulses~\footnote{In Ref.~\cite{Moore1995} a
train of Gaussian pulses was used.}.

Using the rescaled units
 $\phi = 2k_L x$, $p = ({2k_L T}/{M}) P$, and $t = {\tau}/{T}$,
one can write the 
dimensionless Hamiltonian:
\begin{equation}
\tilde{H}= \frac{p^2}{2} + k \cos(\phi) \sum_{{n\in\mathbb{Z}}} f(t - n),
\end{equation}
where $f(r)$ is a dimensionless pulse function of unit amplitude and duration ${\tau_p}/{T}$, $k$ is the dimensionless kick strength given by
$k = ( {8 V_0}/{\hbar}) \omega_r T^2$,
with $\omega_r = {\hbar k_L^2}/{2M}$ being the recoil frequency,
and the rescaled Hamiltonian $\tilde{H}$ is related to the original Hamiltonian 
$H$ via $\tilde{H} = ( {4k_L^2 T^2}/{M}) H$.
In the quantized version of the model, $\phi$ and $p$ are conjugate variables satisfying the commutation relation
$[\phi, p] = i\hbar_{\text{eff}}$,
where the effective Planck constant is defined as
$\hbar_{\text{eff}} = 8 \omega_r T$.

\begin{figure}
\centering
\includegraphics[width=.40\textwidth]{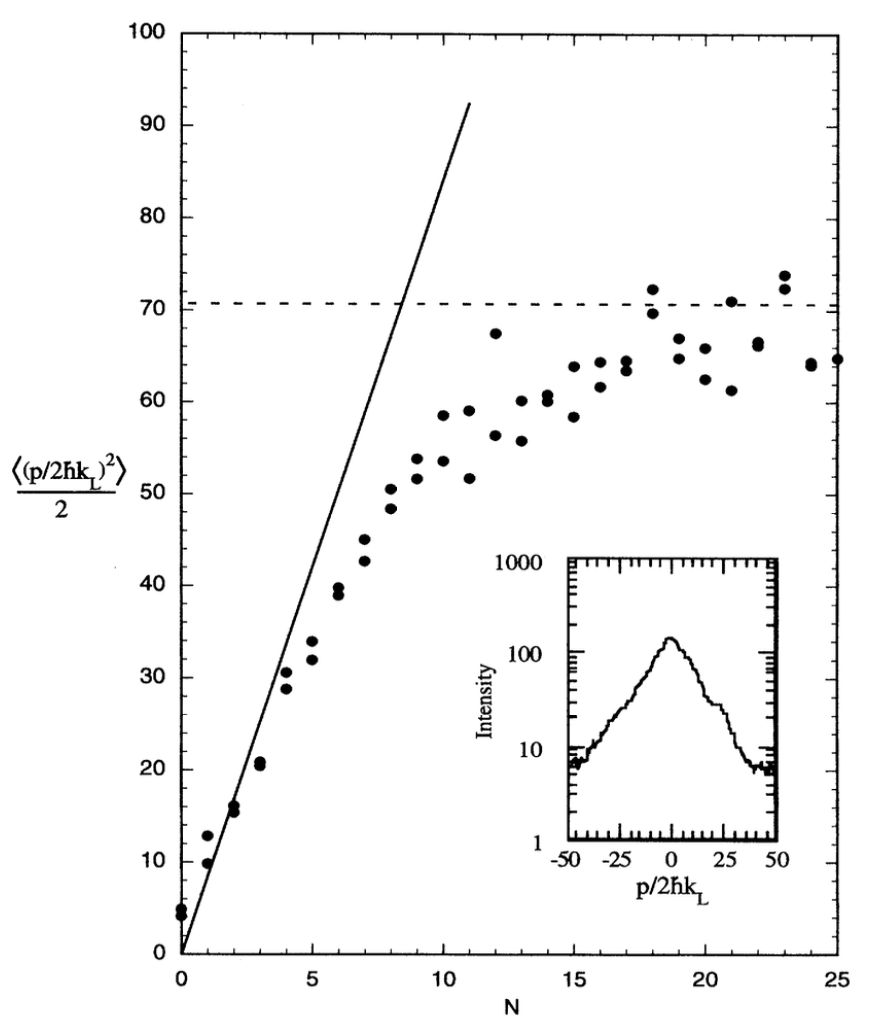}
\caption{Experimental observation of dynamical localization. Solid symbols show the saturation of the mean energy as a function of time $N$ (measured in number of kicks). The dashed line indicates the expected saturation energy from quantum localization theory, while the solid line represents the classical quasi-linear diffusive growth. 
The inset shows the experimentally measured momentum distribution, 
consistent with exponential localization over a length 
$\ell= k^2/2\approx 8.3$.
Figure taken from \cite{Moore1995}. 
}
\label{fig:coldatoms}
\end{figure}

Figure~\ref{fig:coldatoms} provides direct evidence of dynamical localization: the mean energy initially grows linearly with time and then saturates. The inset shows the localized, exponential profile of the momentum distribution for the evolved cold atomic cloud.

\subsubsection{Variants of kicked rotor}

Variants of the kicked-rotor model allow for the experimental investigation of relevant physical phenomena. In this section, we focus on the localization–delocalization transition in a generalized kicked-rotor model, which is analogous to the three-dimensional Anderson model, as well as on quantum computing of a kicked map using actual quantum hardware.

\begin{figure}
\centering
\includegraphics[width=.40\textwidth]{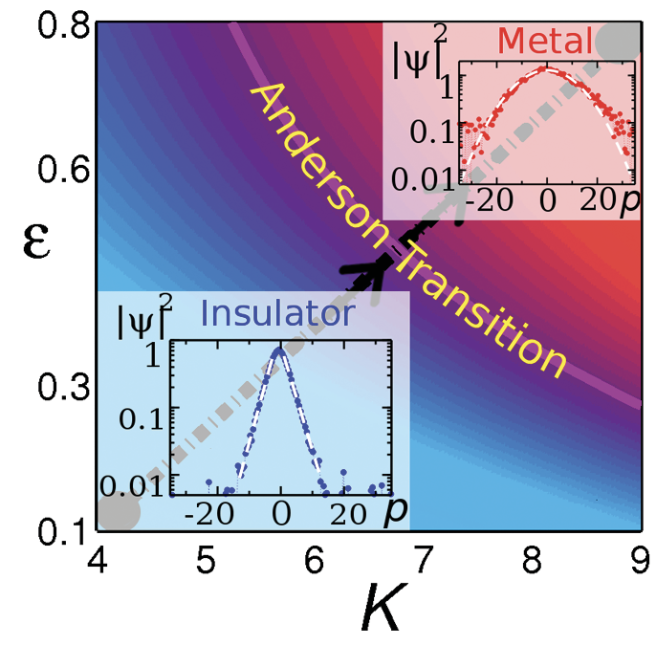}
\caption{Phase diagram of the quasiperiodic kicked rotor obtained from numerical simulations. The localized (insulating) region is shown in blue, while the diffusive (metallic) region appears in red. Experimental parameters are varied along the diagonal dash-dotted line. Insets display the experimentally observed momentum distributions: an exponentially localized profile in the insulating phase and a Gaussian distribution in the diffusive (metallic) phase.
Figure taken from \cite{Chabe2008}. 
}
\label{fig:incommensurate}
\end{figure}

The Anderson metal–insulator transition is a key phenomenon in the study of quantum disordered systems. An insulator is characterized by localized states, whereas a metallic phase generally exhibits diffusive transport, associated with delocalized states.  While all states are localized in one and two dimensions for the standard  Anderson model with uncorrelated disorder, in three dimensions the model predicts a metal–insulator transition: extended and localized states are separated  by a critical energy $E_c$, the mobility edge, and sufficiently strong  disorder localizes all states in the spectrum~\cite{Abrahams1979}. As discussed in Sec.~\ref{sec:Anderson}, there exists a mapping between  the kicked rotor and a one-dimensional tight-binding model with pseudo-random  disorder.  A generalized, \emph{quasiperiodic} kicked-rotor model, 
\begin{equation}
H = \frac{p^2}{2} + K \cos x \left[ 1 + \epsilon \cos(\omega_2 t) \cos(\omega_3 t) \right] \sum_{n} \delta(t - n),
\label{eq:kicked_hamiltonian}
\end{equation}
with the frequencies $\omega_2$ and $\omega_3$ incommensurate with each other, and also with the kicking frequency, 
formally maps ontoan effective three-dimensional Anderson model~\cite{Casati1989} (see also \cite{wang2009} for a three-dimensional generalization of the kicked rotor).
In this model, a transition from an insulating to a metallic phase--that is, from localized to delocalized momentum distributions--can be induced by increasing either the kicking strength $K$ or the amplitude modulation $\epsilon$. These theoretical predictions have been confirmed experimentally using cold cesium atoms exposed to a pulsed laser standing wave~\cite{Chabe2008}, see Fig.~\ref{fig:incommensurate}.

While the experiments described in Sec.~\ref{sec:coldatoms} are 
\textit{quantum simulations} of the kicked-rotor model, 
the dynamics of the same model is also amenable to \textit{quantum computing}, that is, 
it can be numerically evaluated using a universal quantum computer.
In that case the operators $\hat{U}_T$ and $\hat{U}_k$
in (\ref{krotquantum}) can be decomposed into a universal set of 
quantum gates (such as single-qubit and CNOT gates~\cite{qcbook}).
The operators $\hat{U}_T$ and $\hat{U}_k$ are diagonal in the 
action and angle basis, respectively, and one can go from one basis to 
the other by means of the Fourier transform, which can be implemented 
efficiently in a quantum computer~\cite{qcbook}. 
For a kicked rotor with $N$ levels, the Fourier transform 
requires $O((\log_2 N)^2)$ quantum gates, while
$O((\log_2 N)^3)$
quantum gates are required to simulate one map step~\cite{Georgeot2001}, 
whereas the best known classical algorithm, based on the fast Fourier 
transform for the changes of basis, requires 
$O(N\log_2 N)$ operations. Therefore, the quantum algorithm 
which computes the dynamics of the quantum kicked rotor is exponentially
faster than any known classical algorithm. 
Note that efficient implementation of the operator $\hat{U}_k$ necessitates the introduction of auxiliary qubits.
Even more efficient is the quantum computation of the dynamics
of the \emph{quantum sawtooth map},
where a particle moving on a circle is periodically 
kicked as in (\ref{krotham}), but with a potential 
$V(\theta)=-k(\theta-\pi)^2/2$~\footnote{This map is referred to as the sawtooth map since the force $F(\theta) = -\frac{dV(\theta)}{d\theta} = k(\theta - \pi)$ 
has a sawtooth shape, featuring a discontinuity at $\theta = 0$.}.
In this case there is no need for auxiliary qubits and the 
complexity of the quantum algorithm scales as $O((\log_2 N)^2)$~\cite{Benenti2001,Benenti2003}.

\begin{figure}
\centering
\includegraphics[width=.40\textwidth]{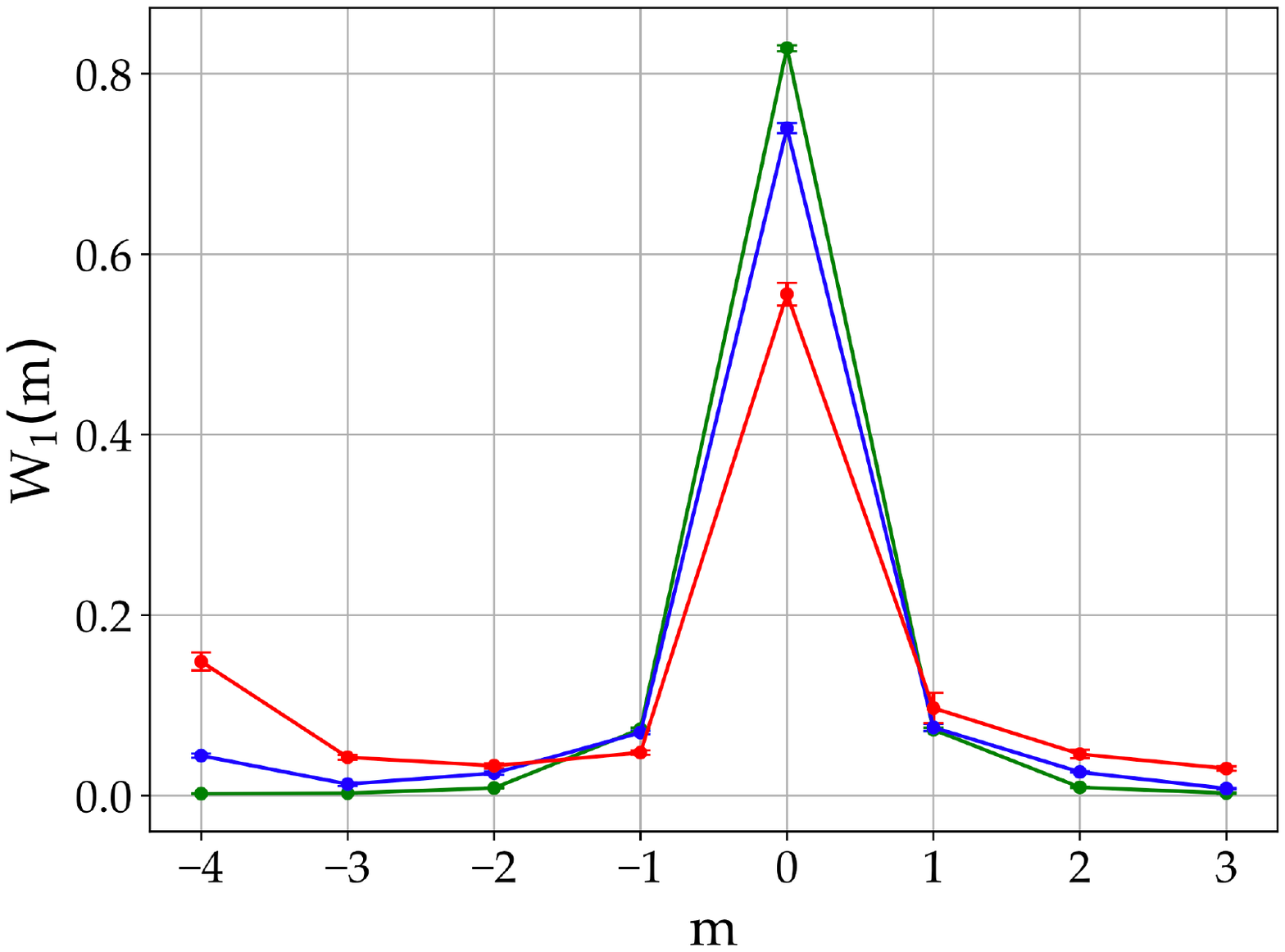}
\caption{Dynamical localization in the quantum sawtooth map with $n=3$ qubits.
The probability distributions obtained from an IBM quantum processor (red), the Qiskit simulator (blue), and an ideal noiseless
simulation (green) are compared after one map step.
Figure adapted from \cite{Pizzamiglio2021}.}
\label{fig:IBM}
\end{figure}

Dynamical localization in the quantum sawtooth map has been observed experimentally on NMR quantum processors~\cite{Henry2006} and, more recently, using a freely available IBM quantum processor with three superconducting qubits, accessed remotely through cloud-based quantum programming~\cite{Pizzamiglio2021}.
In the example shown in Fig.~\ref{fig:IBM}, the initial condition is sharply peaked in action space, with $\psi_0(m) = \delta_{m, m_0}$ ($m_0 = 0$). The kick strength is $k = 0.273 < 1$, so the distribution becomes localized after a single map step. However, $K=kT = 1.5$ corresponds to a regime of diffusive, chaotic behavior in the underlying classical dynamics.
The figure compares the ideal (noiseless) probability distribution after $t = 1$ map step with the results obtained from a real quantum device and from the Qiskit simulator (provided by IBM), which models several relevant noise sources, including dephasing, relaxation, and readout errors.
The results show that the quantum hardware exhibits a localization peak that arises from quantum interference. This is notable because the quantum algorithm involves a forward–backward quantum Fourier transform, effectively exploring the full Hilbert space of the quantum register via a complex, multi-path interferometer. This process leads to wave--function localization--an inherently fragile quantum phenomenon that is highly sensitive to noise.
However, the observed peak height in the hardware results is significantly lower than both the ideal noiseless prediction and the outcome from the Qiskit simulator\footnote{This indicates the existence of noise channels not properly taken into account by the simulator, showing that correctly modeling error sources in a quantum computer is, in itself, a complex problem.}.

\section{Selected advances}\label{sec:advancement}
\subsection{Pseudoclassical theory of near-resonance dynamics}

Although quantum resonances are usually regarded as a purely quantum effect, it turns out that in the near-resonant regime their dynamics can be captured 
by a pseudoclassical (or $\varepsilon$-classical) theory~\cite{Fishman2002,Fishman2003}. The pseudoclassical approach is inspired by a rescaling usually done for the standard semiclassical limit of the quantum kicked rotor, which, as we discussed before, is obtained by simultaneously taking the kick period $T\to 0$ and the kick strength $k\to \infty$, while keeping their product fixed. A similar variable rescaling can be applied for kick periods $T$ that are not small -- and thus far from the standard classical limit -- but lie within the near-resonance region. In what follows, we will first focus on the low-order-resonance cases  -- principal resonance and anti-resonance cases, which correspond to $T=2\pi l$ with even and odd $l$,  respectively --, then show applications to the double kicked-rotor system, and finally present recent extensions to high-order resonances.

For the low-order resonance cases, we consider the situation where the kicking period is slightly detuned from the resonance condition, i.e.,
$T = 2\pi l + \delta$, where $\delta$ is a small irrational multiple of $\pi$ to exclude exact higher-order resonances. For simplicity, we assume $\delta > 0 $ throughout. Then the Floquet operator reads
\begin{equation}
\hat{U} = \exp(-i\pi l  \hat{I}^2)\exp\left(-i\frac{\delta}{2}  \hat{I}^2\right) \exp(-i k\cos\hat{\theta}).\label{eq:delta_EvOp1}
\end{equation}
Using the identity $\exp(-i\pi l  \hat{I}^2) = \exp(-i\pi l  \hat{I})$, the above equation can be expressed as
\begin{equation}
  \hat{U} = \exp(-i\pi l  \hat{I})\exp\left(-i\frac{\delta}{2}  \hat{I}^2\right) \exp(-i k\cos\hat{\theta}).\label{eq:delta_EvOp2}
\end{equation}
In analogy with the semiclassical limit, we now consider taking the limit $\delta \to 0$  and $k \to \infty$, while keeping the product $k\delta \equiv K$ fixed. Note that if we define the rescaled momentum operator $\hat{J} = \delta\hat{I}$, it is straightforward to verify the commutation relation $[\hat{J}, \hat{\theta}] = -i\delta$, which indicates that \( \delta \) plays a role analogous to the Planck constant. Consequently, equation~\eqref{eq:delta_EvOp2} formally defines the quantized version of the Hamiltonian
\begin{equation}
  H(\theta,J;t) = \frac{J^2}{2} + \pi l J + K\cos\theta 
  \sum_{n=-\infty}^{+\infty}
  \delta(t-n).\label{eq:PCham}
\end{equation}
The corresponsing stroboscopic classical map is given by
\begin{equation}
  \left\{
    \begin{aligned}
    \bar{J} &= J + K \sin\theta, \\
    \bar{\theta} &= \theta + \bar{J}+\pi l .\\
    \end{aligned}
  \right.
  \label{eq:PCM}.
\end{equation}
Thus, the pseudoclassical map offers a classical representation of the near-resonant quantum dynamics, with the parameter $\delta$ effectively playing the role of a Planck-like constant. This approach bridges the quantum and classical descriptions, enabling both insightful and tractable analysis of kicked-rotor systems in the vicinity of resonance \cite{Fishman2002,Fishman2003,Abb2009,Sadgrove2011}.

\begin{figure}
\centering
\includegraphics[width=0.40\textwidth]{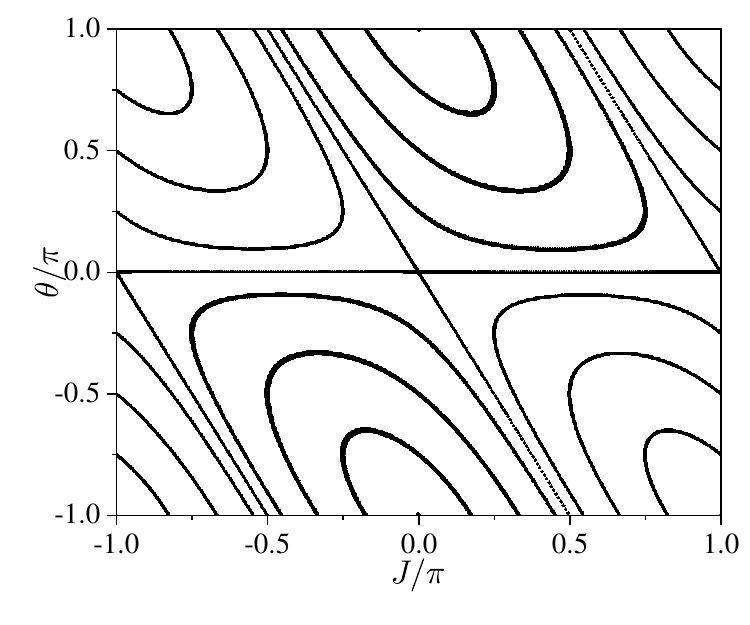}
\caption{Phase-space  portrait of the pseudoclassical map of the double-kicked rotor. Classical kicking strength $K=0.005$.  (Adapted from \cite{Wang2011}.)}
\label{fig:DKphasespace}
\end{figure}

\begin{figure}
\centering
\includegraphics[width=0.40\textwidth]{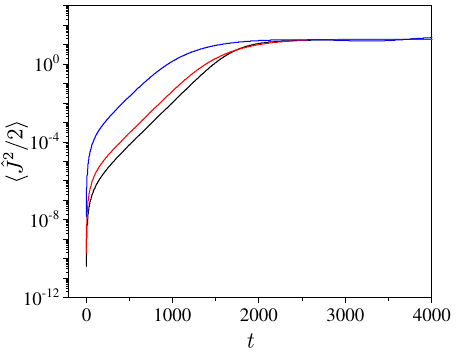}
\caption{Long-lasting exponential quantum spreading in the double kicked rotor. The  blue, red, and black lines correspond to $\delta=10^{-1},10^{-2},10^{-3}$, respectively, with $K=0.005$ fixed.} 
\label{fig:DKexponential}
\end{figure}

A particularly significant application of pseudoclassical theory is its ability to explain the long-lived exponential and superballistic wavepacket diffusion observed in double-kicked rotor systems~\cite{Wang2011}. In this setup, the rotor experiences two kicks within each period $T$, and the corresponding Floquet operator is given by 
\begin{equation}
\hat{U} = \exp\left(-i\frac{(T-T_0)}{2}  \hat{I}^2\right) \exp(-i k\cos\hat{\theta})\exp\left(-i\frac{T_0}{2}  \hat{I}^2\right) \exp(-i k\cos\hat{\theta}).\label{eq:DKEV}
\end{equation}
When $T=2T_0$, the Floquet operator corresponds to two successive steps of the standard kicked rotor with period $T_0$. Interestingly, dynamics emerge when the kicking period $T$ is at a principal resonance and $T_0$ is slightly detuned from specific resonant parameters. Under these conditions, long-time exponential wave-packet spreading can be observed for certain special initial states. A concrete example is given by choosing $T = 4\pi$ and $T_0 = 2\pi + \delta$, where $\delta$ is small. For this set of parameters, the Floquet operator in Eq.~\eqref{eq:DKEV} can be 
seen as the quantized version of the classical map
\begin{equation}
\left\lbrace
  \begin{aligned}
  J' &= J + K\sin\theta,\\
  \theta' &= \theta +\pi + J + K\sin\theta,\\
  \bar{J}&=J' + K\sin\theta',\\
  \bar{\theta}&= \theta' -\pi - J' - K\sin\theta'.\\
  \end{aligned}
\right.\label{eq:DKCL}
\end{equation}
Remarkably, this map can be recognized as a variant of the kicked-Harper model, which possesses nearly straight stable and unstable manifolds in the near-integrable regime \cite{Artuso1992,ARTUSO1994}. 
To see this we perform a change of variables from the classical pair 
$(J,\theta)$ to $(J + \theta,\theta)$,
under which the map in Eq.~\eqref{eq:DKCL} transforms into
\begin{equation}
\left\lbrace
  \begin{aligned}
    \bar{J}+\bar{\theta}&= (J + \theta) + K\sin \theta,\\
  \bar{\theta}&= \theta + K\sin(\bar{J} + \bar{\theta}),\\
  \end{aligned}
\right.\label{eq:KHCL}
\end{equation}
which is precisely the classical map of the kicked-Harper model.  Through linear stability analysis, it can be shown that this model possesses unstable fixed points at $(J + \theta, \theta) = (n\pi, m\pi)$, where $ m, n \in \mathbb{Z}$ and $m + n$ is even. The Lyapunov exponent at each unstable fixed point is given by $\lambda_{\pm} = \ln\lbrack (K^2+2\pm \sqrt{K^4+4K^2})/2\rbrack$.
In the limit $ K \to 0 $, the stable and unstable manifolds of these fixed points form a periodic network, dividing phase-space into cells shaped like parallelograms, with two sides parallel to the $ \theta$-axis and the other two inclined 
at an angle of $\arctan(2)$ (see Fig.~\ref{fig:DKphasespace}). Due to the near-linear structure of these manifolds, classical trajectories in the vicinity of the unstable fixed points can exhibit long-time exponential divergence. 
This phase-space structure provides a compelling explanation for the observed exponential wavepacket spreading. Consider an initial quantum state $|\psi\rangle = |0\rangle$, an eigenstate of  $\hat{J}$ with zero eigenvalue. Classically, this corresponds to an ensemble narrowly distributed along the stable manifold of the unstable fixed point at 
 $(J, \theta) = (0, \pi)$, with an initial width $ \Delta J(0) = \delta $.
Under time evolution, the pseudoclassical map first drives the ensemble toward the fixed point, then repels it exponentially along the unstable manifold at a rate $\lambda_+$. This exponential momentum growth along the unstable manifold continues until 
a time $t$ such that $ J(t) \sim \pi$ (see Fig.~\ref{fig:DKexponential}). 
Using the relation $J(t) \sim \Delta J(0) \exp(\lambda_+ t)$, the time scale for exponential spreading can be estimated as $t_{\rm exp} \sim \lambda_{+}^{-1} \ln(\pi/\delta )$,
which shows excellent agreement with numerical results, as shown in \cite{Wang2011}.

In the above discussion we have restricted ourselves to the kicking potential of the form $k\cos\theta$. If one instead considers other non-KAM singular potentials that satisfy the symmetry $V(\theta)=-V(\theta+\pi)$, for example, $V(\theta) = \frac{2\theta}{\pi} + 1$ for $-\pi \leq \theta < 0$ and $V(\theta) = -\frac{2\theta}{\pi} + 1$ for $0 \leq \theta < \pi$, the energy can exhibit superballistic spreading ($\langle\hat{I}^2\rangle\sim t^3$) which can also be well explained within the pseudoclassical framework. Further details on such systems can be found in \cite{Fang2016}.

Finally, we turn to the case of high-order resonances \cite{Rebuzzini2009,Wang2013a,Zou2024}. When the neighborhood of such a resonance is considered, the Floquet operator with a small detuning $T=4\pi p/q+\delta$ takes the form
\begin{equation}
\hat{U} = \exp\left(-i\frac{2\pi p}{q}\hat{I}^2\right)\exp\left(-i\frac{\delta}{2}\hat{I}^2\right)\exp(-ik\cos\hat{\theta}),
\label{eq:HdeltaEv}
\end{equation}
where $p$ and $q$ are coprime integers. For the last two terms, applying the same rescaling procedure as in the low-order resonance cases yields a well-defined pseudoclassical limit. However, the first term can no longer be simplified to a single translation operator; instead, it decomposes into a superposition of multiple translation operators:
\begin{equation}
\exp\left(-i\frac{2\pi p}{q}\hat{I}^2\right) = \sum_{l=0}^{q-1} G_l \exp\left(-i\frac{2\pi p}{q}l\hat{I}\right),
\end{equation}
where the coefficients $G_l$ are given by a Gaussian sum defined by
\begin{equation}
G_l = \frac{1}{q} \sum_{m=0}^{q-1} \exp\left(-i2\pi\frac{p}{q}m(m-l)\right).
\end{equation}
It can be proven that when $q$ is odd, all $G_l\ne0$, whereas if $q$ is even, exactly half of the $G_l$ coefficients vanish. Physically, this implies that the resonant free evolution can generate $q$ or $q/2$ distinct wavepacket branches, each associated with an angular shift $\frac{2\pi p}{q}l$. By incorporating this with the last two terms in Eq.~\eqref{eq:HdeltaEv}, each quasi-energy band can be effectively approximated by a local pseudoclassical Hamiltonian,
\begin{equation}
H^{(l)}(J,\theta;t)=\frac{J^2}{2}+\frac{2\pi p}{q}l\,J + K\cos\theta\sum_{n=-\infty}^{+\infty}\delta(t-n),
\label{eq:band_H}
\end{equation}
and the full quantum dynamics corresponds to a coherent superposition of contributions from different quasi-energy bands. It now becomes evident that in the high-order near-resonance regime, a single global pseudoclassical approximation based on one Hamiltonian is generally no longer sufficient. Instead, one must adopt a set of local pseudoclassical descriptions, each associated with a distinct quasi-energy band. However, interference between bands can undermine these local approximations and render their predictions unreliable. Consequently, the near-resonant dynamics at high-order resonances cannot be captured by a single classical map alone. Rather, they must be understood as arising from an ensemble of classical maps -- each corresponding to a quasi-energy band -- along with the quantum interference among them. This motivates a formal representation of the pseudoclassical evolution as
\begin{equation}
\mathcal{M}: (J,\theta)\to \{[(J',\theta'+\Delta_j); A_j], j=1,\cdots,\mathcal{N}\}.
\label{eq:HPCM}
\end{equation}
Here $\mathcal{N}=q~\text{or}~q/2$ denotes the number of quasi-energy bands, $(J',\theta')$ is obtained from $(J,\theta)$ via one-step evolution under Hamiltonian $H^{(0)}(J,\theta;t)$ in Eq.~\eqref{eq:band_H}, and each $A_j$ is a nonzero Gaussian sum corresponding to the Bloch phase $\Delta_j$. Compared to the low-order case, in which the classical phase-space evolution $(J(t),\theta(t))$ alone is sufficient, one must now also track the complex amplitudes $\{A_j\}$ associated with different classical branches. When these branches are well separated in phase-space, $|A_j|^2$ can be interpreted as the weight of the $j$-th classical state $(J',\theta'+\Delta_j)$ in evaluating expectation values of observables. However, when different branches overlap, their amplitudes must be combined coherently, and the relative phases become essential for capturing quantum interference. This framework is applicable in principle to arbitrary resonance orders, and its effectiveness has been confirmed by numerical simulations for representative cases \cite{Zou2024}. However, as demonstrated in equation \eqref{eq:HPCM}, for generic high-order resonances, such a framework leads to a rapid proliferation of classical states -- typically growing as $\mathcal{N}^t$, with $t$ the number of map iterations, which limits the method's applicability to short or intermediate time scales. Only certain symmetries of the kicking potential can suppress this proliferation, thereby enabling accurate long-time predictions \cite{Zou2024}.

The pseudoclassical theory has been successfully extended to both non-Hermitian kicked-rotor \cite{Zou2024} and kicked-top systems \cite{Zou2022,Zou2025}. In the non-Hermitian kicked rotor, the intrinsic dissipative dynamics effectively suppresses the exponential proliferation of classical states, thereby addressing the major challenge of state growth for high-order resonances \cite{Zou2024}. In the case of the kicked top system, the pseudoclassical framework accurately predicts wave packet recurrence and offers a comprehensive explanation of the dynamics of entanglement entropy \cite{Zou2022}. Moreover, it may lead to promising applications in quantum metrology by enabling enhanced precision through near-resonant dynamics \cite{Zou2025}. 

\subsection{Floquet topological phases}

Periodic driving has become a foundational technique for engineering topological phases in contemporary condensed matter physics. The kicked rotor, as a paradigmatic Floquet system, exhibits a rich landscape of Floquet topological phases. Particularly intriguing is the resonant regime, where the kicking period matches specific system frequencies, leading to an effective Floquet Hamiltonian that maps onto a tight-binding model in momentum space. This emergent lattice structure enables the realization of topological phases, with internal degrees of freedom playing a pivotal role in their formation. A variety of exotic Floquet topological phases have been theoretically proposed within this framework. In the following, we will first introduce the topological phases arising in the double kicked rotor, as well as its quantized adiabatic transport \cite{Ho2012}, and subsequently present how the spin-1/2 double kicked rotor can be employed to realize all nontrivial one-dimensional topological phases \cite{Zhou2018,Koyama2023}.

To characterize the band topology of the on-resonance double-kicked rotor, we introduce an additional
periodic phase parameter $\alpha\in(0,2\pi] $, which plays a role analogous to that of a synthetic dimension. The Floquet operator is given by
\begin{equation}
   \hat{U}(\alpha) = \exp\left(-i\frac{(T-T_0)}{2}  \hat{I}^2\right) \exp(-i k\cos\hat{\theta})\exp\left(-i\frac{T_0}{2}  \hat{I}^2\right) \exp(-i k\cos(\hat{\theta}+\alpha)).
\end{equation}
Here we set $T=4\pi$ and $T_0 = 2\pi p/q$. Under these conditions, $\hat U(\alpha)$ is perfectly periodic in momentum space with a period of $q$, which allows for the application of Bloch's theorem, and the Floquet eigenvalue problem
\begin{equation}
\hat U(\alpha) |\psi_n(\phi,\alpha)\rangle= e^{i\omega_n(\phi,\alpha)} |\psi_n(\phi,\alpha)\rangle
\end{equation}
defines $q$ Floquet bands $\omega_n(\phi,\alpha)$, where $\phi\in[0,2\pi)$ is the Bloch phase in momentum space. The eigenstates can be factorized as $|\psi_n(\phi,\alpha)\rangle = \hat{X}(\phi) |u_n(\phi,\alpha)\rangle$, with $\langle m|u_n(\phi,\alpha)\rangle =\langle m +lq |u_n(\phi,\alpha)\rangle$ and $\hat{X}(\phi)\equiv e^{i\hat{I}\phi/q}$. To obtain a reduced representation, we define the projected state $|{\bar u_n(\phi,\alpha)}\rangle = \sum_{m=1}^q |m\rangle\langle m| u_n(\phi,\alpha)\rangle$. It can be easily shown that
\begin{equation}
\hat{\bar U}(\phi,\alpha) |{\bar u_n(\phi,\alpha)}\rangle = e^{i\omega_n(\phi,\alpha)}|\bar u_n(\phi,\alpha)\rangle,
\label{eq:eigUbar}
\end{equation}
where $\hat{\bar U}(\phi,\alpha)$ is a $q\times q$ Floquet matrix defined by
\begin{equation}
\langle m| \hat{\bar U}(\phi,\alpha)|m'\rangle =
\sum_{l=-\infty}^{\infty}
\langle m|\hat X^\dagger(\phi)\hat U(\alpha)\hat X(\phi) |m' + lq\rangle.
\label{eq:Ubar}
\end{equation}

To characterize the topology of these bands, we consider the two-torus defined by $(\phi,\alpha)$. For each band $n$, the Berry connection is defined as
\begin{equation}
A_{n,\mu}(\phi,\alpha) = i\langle \bar u_n(\phi,\alpha)|\partial_\mu|\bar u_n(\phi,\alpha)\rangle,\qquad \mu\in\{\phi,\alpha\},
\end{equation}
and the corresponding Berry curvature is
\begin{equation}
\begin{split}
B_n(\phi,\alpha) &= \partial_\phi A_{n,\alpha}-\partial_\alpha A_{n,\phi} \\
&= i\left[\left(\frac{\partial}{\partial\phi}\langle\bar u_n(\phi,\alpha)|\right)\frac{\partial}{\partial\alpha}|\bar u_n(\phi,\alpha)\rangle-\left(\frac{\partial}{\partial\alpha}\langle\bar u_n(\phi,\alpha)|\right)\frac{\partial}{\partial\phi}|\bar u_n(\phi,\alpha)\rangle\right]\\
&= i\sum_{\substack{n'=1\\n'\neq n}}^{q}\left[\left(\frac{\partial}{\partial\phi}\langle\bar u_n(\phi,\alpha)|\right)|\bar u_{n'}(\phi,\alpha)\rangle\langle\bar u_{n'}(\phi,\alpha)|\frac{\partial}{\partial\alpha}|\bar u_n(\phi,\alpha)\rangle-\left(\frac{\partial}{\partial\alpha}\langle\bar u_n(\phi,\alpha)|\right)|\bar u_{n'}(\phi,\alpha)\rangle\langle\bar u_{n'}(\phi,\alpha)|\frac{\partial}{\partial\phi}|\bar u_n(\phi,\alpha)\rangle\right].
\end{split}
\label{eq:Berry_curv}
\end{equation}
Physically, $B_n(\phi,\alpha)$ captures the local geometric phase accumulation of the Floquet eigenstates over the parameter space. The topological invariant associated with each band is the Chern number, defined by integrating the Berry curvature over the Brillouin-zone torus:
\begin{equation}
C_n = \frac{1}{2\pi}\int_0^{2\pi}\!\!\int_0^{2\pi} B_n(\phi,\alpha)\,d\phi\,d\alpha.
\label{eq:Chern}
\end{equation}
This integer quantity classifies the global topology of the Floquet bands and is robust against any smooth deformations of system parameters that preserve the quasi-energy gaps.

To efficiently compute the Berry curvature in terms of the Floquet operator and its derivatives, we differentiate Eq.~\eqref{eq:eigUbar} with respect to $\phi$, and take the inner product with $\langle \bar{u}_{n'}(\phi, \alpha) | $, and rearrange to obtain the Feynman-Hellmann-like relation as
\begin{equation}
  \left\langle \bar{u}_{n'}(\phi, \alpha) \right| \frac{\partial}{\partial \phi} \left| \bar{u}_{n}(\phi, \alpha) \right\rangle = \frac{\left\langle \bar{u}_{n'}(\phi, \alpha) \middle| \frac{\partial \bar{U}(\phi, \alpha)}{\partial \phi} \middle| \bar{u}_{n}(\phi, \alpha) \right\rangle}{e^{i \omega_{n}} - e^{i \omega_{n'}}},\label{eq:parphi}
\end{equation}
which has adjoint given by
\begin{equation}
  \left(\frac{\partial}{\partial \phi}\langle\bar{u}_{n}(\phi, \alpha)|\right)|\bar{u}_{n'}(\phi, \alpha)\rangle = \frac{\langle\bar{u}_{n}(\phi, \alpha)|\frac{\partial \bar{U}^{\dagger}(\phi, \alpha)}{\partial \phi}|\bar{u}_{n'}(\phi, \alpha)\rangle}{e^{-i\omega_{n}}-e^{-i\omega_{n'}}}.\label{eq:adjparphi}
\end{equation}
Similar expressions also hold for the $\frac{\partial}{\partial \alpha}$ derivative. Substituting Eqs.~\eqref{eq:parphi}, \eqref{eq:adjparphi} and the corresponding results for $\frac{\partial}{\partial \alpha}$ into Eq.~\eqref{eq:Berry_curv}, we obtain a convenient expression for the Berry curvature,
\begin{equation}
  B_{n}(\phi, \alpha) = i \sum_{\substack{n'=1 \\ n'\neq n}}^{q} \left\{ \frac{\langle \bar{u}_{n}(\phi,\alpha) | \frac{\partial \bar{U}^{\dagger}}{\partial \phi} | \bar{u}_{n'}(\phi,\alpha)\rangle \langle \bar{u}_{n'}(\phi,\alpha) | \frac{\partial \bar{U}}{\partial \alpha} | \bar{u}_{n}(\phi,\alpha)\rangle}{|e^{i\omega_{n}} - e^{i\omega_{n' }}|^{2}} - \mathrm{c.c.} \right\}. \label{eq:res_Berry_curv}
\end{equation}
This form is particularly suitable for numerical evaluation, since it only involves matrix elements of the derivatives of the finite-dimensional Floquet operator, with $\partial_\phi\bar{U}^\dagger$ paired with $\partial_\alpha\bar{U}$ in the numerator.


\begin{figure}
\centering
\includegraphics[width=0.55\textwidth]{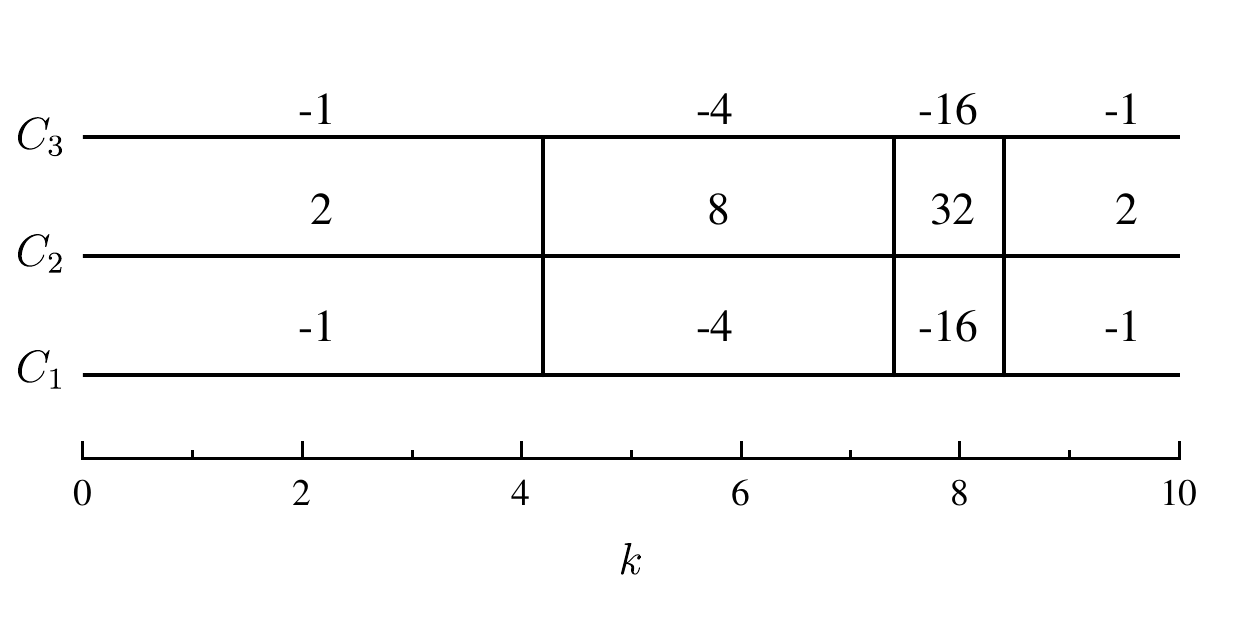}
\caption{  The Chern numbers of the double kicked rotor for $T_0 = 2\pi/3$. (Adapted from \cite{Ho2012}.)}
\label{fig:topology_chern}
\end{figure}

The results of the Chern numbers are shown in Fig. \ref{fig:topology_chern}. It can be seen that at certain isolated critical values of $k$ the Chern numbers undergo jumps, signaling the presence of topological phase transitions in our double-kicked rotor model. Indeed, the Chern numbers are integer--valued invariants under smooth deformations of the bands and change only discontinuously when bands touch.

A key physical consequence of a nonzero Chern number is the quantization of momentum transport when the synthetic dimension parameter $\alpha$ is slowly cycled. To begin, we construct the initial state as follows:
\begin{equation}
  |\Psi_n\rangle = \frac{1}{2\pi} \int_0^{2\pi}d \phi\,| \psi_n(\phi,\alpha=0)\rangle,
\end{equation}
which represents an equal-weight superposition of all the Floquet eigenstates of band $n$. This coherent superposition state, which can be interpreted as a Wannier function in momentum space, uniformly samples all the Bloch eigenstates with different values of $\phi$, with a profile localized in the momentum space. Next, we consider an adiabatic evolution of the parameter $\alpha$ via a discretized protocol $\alpha_d = 2\pi d/d_f$ for $(d-1)T\le t<d T$, such that a full adiabatic cycle is completed at $t=d_f T$. Under the adiabatic approximation, the system remains in a superposition of instantaneous eigenstates of the Floquet operator $\hat{U}(\alpha)$. The time-evolved state at $t=(dT)^-$ is then given by
\begin{equation}
  |\Psi_n(t)\rangle = \frac{1}{2\pi} \int_0^{2\pi}d \phi\,| \psi_n(\phi,\alpha_d)\rangle e^{i\theta(\phi,\alpha_d)},
\end{equation}
where $\theta(\phi,\alpha_d)$ denotes the accumulated dynamical and geometric phases.

The central quantity of interest is the expectation value of the momentum operator $\hat{I}$. We show that the calculation simplifies remarkably, with the final momentum expectation change being purely geometric in origin. We first evaluate the expectation value of the momentum operator $\hat{I}$ in this state. According to the definition, it is expressed as
\begin{equation}
  \langle I_d\rangle  = \sum_{m=-\infty}^{\infty} m  \int_{0}^{2\pi} d\phi \int_{0}^{2\pi} d\phi' e^{\frac{i \, m(\phi - \phi')}{q}} \langle m | u_{n}(\phi, \alpha_{d}) \rangle \langle u_{n}(\phi', \alpha_{d}) | m \rangle \times \frac{1}{4\pi^{2}} e^{i(\theta(\phi,\alpha_d) - \theta(\phi',\alpha_d))}.
\end{equation}
Using the identity $m e^{im(\phi -\phi')/q}=-i q\frac{\partial }{\partial \phi}e^{i m(\phi -\phi')/q}$, we integrate by parts and decompose $m$ into $m = lq+m'$, where $l\in \mathbb{Z}$ and $m'\in\{1,2,\cdots,q\}$. This yields:
\begin{equation}
\begin{split}
  \langle I_d\rangle  &= \frac{i q}{4\pi^2}\sum_{m'=1}^{q}\sum_{l=-\infty}^{\infty} \int_{0}^{2\pi} d\phi \int_{0}^{2\pi} d\phi' e^{\frac{i \, (m'+lq)(\phi - \phi')}{q}} \frac{\partial}{\partial \phi}\left( \langle m'+lq | u_{n}(\phi, \alpha_{d}) \rangle \langle u_{n}(\phi', \alpha_{d}) | m'+lq \rangle e^{i(\theta(\phi,\alpha_d) - \theta(\phi',\alpha_d))}\right)\\
  &= \frac{i q}{4\pi^2}\sum_{m'=1}^{q} \int_{0}^{2\pi} d\phi \int_{0}^{2\pi} d\phi' e^{\frac{i \, m'(\phi - \phi')}{q}}\sum_{l=-\infty}^{\infty} e^{i l(\phi - \phi') } \left\lbrack\left(\frac{\partial}{\partial \phi} \langle m' | u_{n}(\phi, \alpha_{d}) \rangle\right) + i\frac{\partial\theta(\phi,\alpha_d)}{\partial \phi}\langle m' | u_{n}(\phi, \alpha_{d}) \rangle \right\rbrack \langle u_{n}(\phi', \alpha_{d}) | m' \rangle e^{i(\theta(\phi,\alpha_d) - \theta(\phi',\alpha_d))}.
\end{split}
\end{equation}
Applying the Poisson summation formula, $\sum_{l=-\infty}^{\infty} e^{il(\phi - \phi') } = 2\pi\delta_{2\pi}(\phi-\phi') $, we arrive at
\begin{equation}
  \langle I_d\rangle  = \frac{i q}{2\pi} \int_{0}^{2\pi} d\phi \sum_{m'=1}^{q}\left(\frac{\partial}{\partial \phi} \langle m' | u_{n}(\phi, \alpha_{d}) \rangle\right) \langle u_{n}(\phi, \alpha_{d}) | m' \rangle - \frac{q}{2\pi} \int_{0}^{2\pi} d\phi \frac{\partial\theta(\phi,\alpha_d)}{\partial \phi}\sum_{m'=1}^{q}|\langle m' | u_{n}(\phi, \alpha_{d}) \rangle|^2.
\end{equation}
The second term vanishes because $\sum_{m'=1}^{q}|\langle m'|u_n(\phi,\alpha_d)\rangle|^2=1$ and, in the chosen periodic gauge, $\theta(2\pi,\alpha_d)=\theta(0,\alpha_d)$. The momentum expectation simplifies to a compact form:
\begin{equation}
  \langle I_d\rangle = \frac{1}{2\pi} \int_{0}^{2\pi} d\phi \langle \bar{u}_{n}(\phi, \alpha_{d})| iq \frac{\partial}{\partial \phi} |\bar{u}_{n}(\phi, \alpha_{d}) \rangle .\label{eq:avep}
\end{equation}
Here, the identity $\langle \bar{u}_{n}(\phi, \alpha_{d})| \frac{\partial}{\partial \phi} |\bar{u}_{n}(\phi, \alpha_{d}) \rangle = \sum_{m'=1}^{q}\left(\frac{\partial}{\partial \phi} \langle m' | u_{n}(\phi, \alpha_{d}) \rangle\right) \langle u_{n}(\phi, \alpha_{d}) | m' \rangle$ follows straightforwardly from the definition of $|\bar{u}_{n}(\phi, \alpha) \rangle$.

By considering the infinitesimal change in momentum over one adiabatic step, we can directly link the transport to the Berry curvature. To show this, we first examine the incremental change $\langle I_{d}\rangle - \langle I_{d-1}\rangle$. Using first-order perturbation theory, the Floquet eigenstate with parameter $\alpha_d$ can be expanded as
\begin{equation}
|\bar u_n(\phi,\alpha_{d})\rangle = |\bar u_n(\phi,\alpha_{d-1})\rangle  +  |\bar u_n^{(1)}(\phi,\alpha_{d-1})\rangle,
\label{eq:first_perturb}
\end{equation}
where
\begin{equation}
 |\bar u_n^{(1)}(\phi,\alpha_{d-1})\rangle =\delta\alpha \sum_{n'\neq n}
\frac{\langle \bar u_{n'}(\phi,\alpha_{d-1})|\partial_{\alpha_{d-1}}\hat{\bar U}|\bar u_n(\phi,\alpha_{d-1})\rangle}{e^{i\omega_n}-e^{i\omega_{n'}}}|\bar u_{n'}(\phi,\alpha_{d-1})\rangle.\label{eq:perturb}
\end{equation}
Using Eqs.~\eqref{eq:avep} and \eqref{eq:first_perturb} and retaining terms linear in $\delta\alpha$, the momentum change can be expressed as
\begin{equation}
\begin{split}
\langle I_{d}\rangle - \langle I_{d-1}\rangle &= \frac{iq}{2\pi} \int_{0}^{2\pi} d\phi  \left\lbrack \langle\bar u_n^{(1)}(\phi,\alpha_{d-1})|  \frac{\partial}{\partial \phi} |\bar{u}_{n}(\phi, \alpha_{d-1}) \rangle  + \langle\bar u_n(\phi,\alpha_{d-1})|  \frac{\partial}{\partial \phi} |\bar{u}_{n}^{(1)}(\phi, \alpha_{d-1}) \rangle\right\rbrack\\
&= \frac{iq}{2\pi} \int_{0}^{2\pi} d\phi  \left\lbrack \langle\bar u_n^{(1)}(\phi,\alpha_{d-1})|  \frac{\partial}{\partial \phi} |\bar{u}_{n}(\phi, \alpha_{d-1}) \rangle  - \left( \frac{\partial}{\partial \phi} \langle\bar u_n(\phi,\alpha_{d-1})| \right) |\bar{u}_{n}^{(1)}(\phi, \alpha_{d-1}) \rangle\right\rbrack.\\
\end{split}
\label{eq:deltape}
\end{equation}
Substituting the explicit form of $ |\bar u_n^{(1)}(\phi,\alpha_{d-1})\rangle$ from Eq.~\eqref{eq:perturb} and using the relations in Eqs.~\eqref{eq:parphi} and \eqref{eq:adjparphi}, we finally obtain
\begin{equation}
\langle I_{d}\rangle - \langle I_{d-1}\rangle = -\frac{q}{2\pi} \int_{0}^{2\pi} d\phi B_{n}(\phi,\alpha_{d-1}) \delta\alpha,
\end{equation}
where $B_{n}(\phi,\alpha)$ is exactly the Berry curvature we obtained in Eq.~\eqref{eq:res_Berry_curv}. Summing over a full adiabatic cycle, the total momentum expectation change is
\begin{equation}
  \delta I = -\frac{q}{2\pi} \int_{0}^{2\pi} d\phi \int_{0}^{2\pi} d\alpha B_{n}(\phi, \alpha)\delta\alpha = -q C_{n},
\end{equation}
where $C_n$ is the Chern number of the $n$th Floquet band. This elegant relation shows that each complete cycle in $\alpha$ pumps an integer number of momentum quanta, proportional to the Chern number of the occupied Floquet band. Hence, the double kicked rotor realizes a topological Thouless pump in momentum space.

\begin{figure}
\centering
\includegraphics[width=0.55\textwidth]{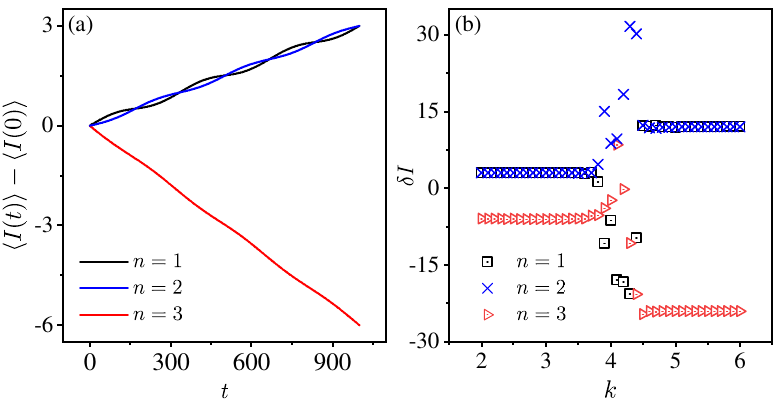}
\caption{(a) Time evolution of the average momentum during one adiabatic cycle for $k=2.0$. (b) Change in the momentum expectation value after one adiabatic cycle implemented in $d_f = 1000$ discretized steps for different $k$. Here $T_0 = 2\pi/3$. (Adapted from \cite{Ho2012}.)}
\label{fig:topology_pumping}
\end{figure}

Detailed results are shown in Fig.~\ref{fig:topology_pumping}, which almost perfectly match the analysis, except near the breakdown around some critical value of $k$. This deviation arises because, when a topological phase transition is about to occur, the associated band gaps become too small to guarantee adiabaticity, leading to imperfect quantization of the pumping process.

It is worth mentioning that the topological phenomena observed in the above double-kicked rotor are deeply rooted in its mathematical equivalence to the kicked Harper model~\cite{Wang2013}, whose quasi-energy spectrum possesses a topologically non-trivial fractal structure characterized by nonzero Chern numbers associated with each spectral band.

Building upon the success of the standard double kicked rotor model in revealing non-equilibrium topological phases, a powerful generalization is the spin-1/2 double kicked rotor \cite{Zhou2018,Koyama2023}. 
To explore a broader landscape of topological phases, the model incorporates an internal spin-$1/2$ degree of freedom, allowing for both spin-dependent and spin-independent kicks, as described by the Floquet operator
\begin{equation}
  \hat{U} = \exp\left(i\frac{\pi}{2}\hat{I}^2\otimes \hat{\sigma}_0\right)\exp\left(-i\hat{H_2}\right)\exp\left(-i\frac{\pi}{2}\hat{I}^2\otimes \hat{\sigma}_0\right)\exp\left(-i \hat{H_1}\right),
\end{equation}
where each kick Hamiltonian $\hat{H}_j$ $(j=1,2)$  is given by
\begin{equation}
\hat{H}_j = K_j^0 \cos\left(\nu_j^0\hat{\theta} + \alpha_j^0\right)\otimes \hat{\sigma}_0 + K_j \cos\left(\nu_j\hat{\theta} + \alpha_j\right)\otimes \bm{n_j\cdot\hat{\sigma}}.
\end{equation}
Here, $\bm{n}_j$ is a unit vector, $\hat{\sigma}_0$ is the identity operator in spin space, and $\hat{\bm{\sigma}}=(\hat{\sigma}_x,\hat{\sigma}_y,\hat{\sigma}_z)$ is the vector of Pauli matrices. Note that the system is at resonance, the Floquet operator is block diagonal in terms of the ``Bloch wave number" $\phi$. Therefore, it is convenient to describe the symmetry properties in terms of the Floquet operator $\hat{U}(\phi)$. According to the Altland-Zirnbauer (AZ) classification, the  time-reversal symmetry, the particle-hole symmetry, and the chiral symmetry for the Floquet operator are respectively given by \cite{Roy2017}:
\begin{equation}
  \hat{T}\hat{U}(\phi) \hat{T}^{-1} =  \hat{U}^{-1}(-\phi), \quad \hat{C}\hat{U}(\phi) \hat{C}^{-1} =  \hat{U}(-\phi),\qquad \hat{\Gamma}\hat{U}(\phi) \hat{\Gamma}^{-1} =  \hat{U}^{-1}(\phi),
\end{equation}
where $\hat{T}$, $\hat{C}$ and $\hat{\Gamma}$ are  time-reversal, particle-hole, and chiral operators, respectively. Based on these symmetries,  nontrivial one-dimensional topological phases can belong to several AZ classes~\cite{Roy2017}. Crucially, it has been demonstrated that, by appropriately choosing the kicking parameters $\nu_j^0,\alpha_j^0$ and $\nu_j,\alpha_j$, the spin-$1/2$ double-kicked rotor can realize all five Altland--Zirnbauer symmetry classes supporting nontrivial topological phases in one dimension, namely AIII, BDI, D, DIII, and CII~\cite{Koyama2023}. For example, the CII class (with $\hat T^2=\hat C^2=-1$) can be realized when all kicking frequencies $\nu_{1,2}^0,\nu_{1,2}$ are odd integers, and the lattice phases are chosen to satisfy $\alpha_1^0=0$, $\alpha_1=\pi/2$, $\alpha_2^0=\pi/2$, and $\alpha_2=0$ \cite{Koyama2023}.

The rich topological structure exhibited by the kicked rotor and its variants establishes it as a robust, non-equilibrium platform for a wide range of physical investigations. Recent advances have successfully extended these studies to incorporate effects such as non-Hermitian physics~\cite{Zhou2019} and the emergence of higher-order topological phases~\cite{Zhou2021} in extended kicked rotor systems, highlighting their capability to model complex and frontier phenomena in condensed matter physics. 

\subsection{Scaling theory and Anderson transition}

In Sec.~\ref{sec:Anderson}, we established that dynamical localization in the kicked rotor can be mapped onto Anderson localization. The one-parameter scaling theory provides a powerful framework for analyzing localization phenomena and phase transitions in the standard Anderson model. According to this theory, disordered systems can be classified into three distinct regimes: the localized phase, where wave functions are exponentially confined and transport is strongly suppressed; the metallic phase, characterized by extended states and diffusive transport; and the critical phase at the metal–insulator transition, where the system displays scale-invariant behavior and multifractal eigenstates.
In what follows, we briefly introduce the one-parameter scaling theory~\cite{Abrahams1979} and then discuss how it can be adapted to the kicked rotor model~\cite{GarciaGarcia2008, wang2009}. Finally, we summarize some recent advances in exploring the Anderson transition within this framework~\cite{Chen2023,Chen2024,Chen2024a}.

\begin{figure}
\centering
\includegraphics[width=.45\textwidth]{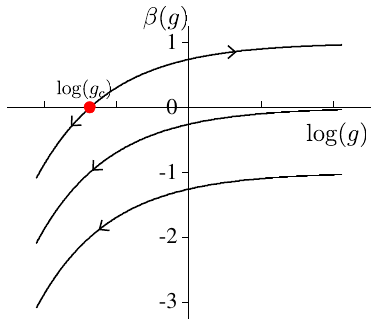}
\caption{Schematic plot of the scaling function $\beta$ as a function of $\log(g)$  in different spatial dimensions ($d=1,2,3$, shown from bottom to top). }
\label{fig:dimen_conduct}
\end{figure}

The core concept of one-parameter scaling theory is the dimensionless conductance $g$, whose scaling behavior is captured by the function $\beta(g)$, introduced by Edwards and Thouless~\cite{Edwards1972}. This conductance is defined as the ratio of the Thouless energy $E_c$ to the mean level spacing $\Delta$, which typically scales as $\Delta\propto 1/L^d$, where $L$ is the system's linear size and $d$ its spatial dimensionality. The Thouless energy characterizes the energy scale associated with the time required for a particle to diffuse across the sample. In the semiclassical limit, it can be expressed as $E_c = \hbar D_{\rm cl}/L^2$ where $D_{\rm cl}$ is the classical diffusion constant. Combining these expressions yields the scaling relation $g\propto L^{d-2} $ for diffusive systems. In contrast, when quantum interference leads to exponential localization of the wave function, the conductance behaves as $g\propto \exp\left(-L/\xi\right)$ where $\xi$ is the localization length. Thus qualitatively, in the thermodynamic limit $L\to\infty$,  the dimensionless conductance diverges ($g\to \infty$) in the metallic phase, while it vanishes ($g\to 0$) in the insulating phase. The dependence of $g$ on the system size therefore serves as a diagnostic for localization. To formalize this, one defines the scaling function $\beta(g) = \partial \log(g(L))/\partial \log(L)$. From the above, we find $\beta(g) = d-2$ for a classical metal (neglecting quantum corrections), and $\beta(g) =\log(g)<0$ for an insulator. 
Assuming that $\beta(g)$ is a continuous and monotonic function, and that the variation of the conductance with system size depends only on the conductance itself, we can analyze how localization corrections depend on the spatial dimensionality $d$. When $d=1$, for $g\to\infty$, $\beta(g)\to -1<0$ and for $g\to 0$, $\beta(g)=\log(g)<0$. Therefore $\beta(g)<0$ for any $g$, and the dimensionless conductance always decreases with the system size and the system will be an insulator in the limit $L\to \infty$ for any nonzero disorder strength. When $d=2$, for $g\to\infty$, $\beta(g)\to0$ and for $g\to 0$, $\beta(g)=\log(g)<0$. We have to determine whether $\beta(g)$ is positive or negative (or it remains zero) once quantum interference effects are included. When $d=3$, for $g\to\infty$, $\beta(g)\to 1 $ and for $g\to 0$, $\beta(g)=\log(g)<0$. Since $\beta(g)$ is continuous and monotonic there must be $g = g_c$ such that $\beta(g_c) = 0$. At $g = g_c$ the system undergoes a metal-insulator transition. All these analyses are summarized in the schematic plot shown in Fig.~\ref{fig:dimen_conduct}.

Now let us see how the one-parameter scaling theory can be adapted to the context of the kicked rotor. In this system, the mean level spacing $\Delta$ is well defined; the main challenge lies in defining the Thouless energy. In the Anderson model, the Thouless energy is typically estimated after averaging over many disorder realizations. 
Since localization in the kicked rotor is defined in momentum space, our task is to determine a typical and meaningful travel time associated with the exploration of the momentum basis. Intuitively, the ensemble average in the Anderson model can be replaced by an average over initial conditions in the kicked rotor. However, in mixed systems, the result depends sensitively on the choice of initial conditions: if the particle is initially located in an integrable region, the typical diffusion time will be larger than in a chaotic region, since KAM barriers slow down transport in momentum space. Therefore, to properly define the dimensionless conductance in the kicked rotor, we must impose the condition that the classical phase-space is homogeneous. Under this assumption, a characteristic travel time can be defined unambiguously.

We are now ready to define $g$ and construct the one-parameter scaling theory for the kicked rotor. Because the classical dynamics of this system is not always purely diffusive, we consider a generalized diffusion law, $\langle \hat{I}^2\rangle\sim t^\mu$ with $\mu>0$. To define finite-size scaling, we consider a family of truncated momentum-space systems with linear extent $N$ in each of the $d$ effective dimensions, so that the Hilbert-space dimension scales as $N^d$. Here $d$ denotes the effective dimension of the momentum-space lattice, rather than necessarily the physical spatial dimension. The Thouless energy is then given by $E_c\propto N^{-2/\mu}$. For the mean level spacing, we consider the generalized scaling form $\Delta\propto N^{-d/d_e}$, which also allows for anomalous spectral scaling associated with singular continuous spectra. Here $d_e$ is the spectral dimension characterizing the local scaling of the spectral measure, such that the number of levels in a sufficiently small energy interval $\delta E$ scales as $N^d(\delta E)^{d_e}$. For a regular spectrum, one has $d_e=1$, whereas certain fractal spectra are characterized by $0<d_e<1$. The dimensionless conductance then scales as
\begin{equation}
  g(N)=\frac{E_c}{\Delta}=N^{\gamma_{\rm cl}},\qquad\gamma_{\rm cl} = \frac{d}{d_e} - \frac{2}{\mu}.
\end{equation}
In order to study how quantum effects modify the dimensionless conductance with increasing system size, we can define the scaling function 
\begin{equation}
  \beta(g) = \frac{\partial \log(g(N))}{\partial \log(N)}.
\end{equation}
In cases where quantum effects are negligible, we have
\begin{equation}
  \beta (g) = \gamma_{\rm cl} = \frac{d}{d_e} - \frac{2}{\mu}.\label{eq:scal_fun}
\end{equation}
Under the assumptions of the one-parameter scaling theory and using Eq.~\eqref{eq:scal_fun}, we can propose the following classification of possible localization regimes in kicked rotor systems: (i) If $\gamma_{\rm cl} > 0$, the system favors a metallic phase with delocalized eigenfunctions and Wigner--Dyson spectral statistics; (ii) If $\gamma_{\rm cl} <0$, the system favors an insulating phase with localized eigenfunctions and Poisson spectral statistics; (iii) If $\gamma_{\rm cl} =0$, the semiclassical conductance becomes scale independent, suggesting the possibility of a metal--insulator transition characterized by multifractal eigenfunctions and critical spectral statistics. These semiclassical predictions may be further modified by quantum interference effects. When the spectral dimension is normal, $d_e=1$, the one-parameter scaling theory predicts that if the classical motion is such that $\mu< \frac{2}{d}$ the quantum properties are those of an insulator, regardless of whether the classical motion is chaotic. For instance, in $d=3$, any classical motion with $\mu < \frac{2}{3}$ is expected to drive the quantum system into the insulating phase---even if the classical dynamics is fully chaotic. Another illustrative example is the standard one-dimensional kicked rotor with a smooth potential. Although the classical dynamics typically exhibits normal diffusion $\mu = 1$, the resulting exponent $\gamma_{\rm cl}=-1<0$ implies that the system resides in the insulating phase. Notably, Eq.~\eqref{eq:scal_fun} also suggests the possibility of a metal--insulator transition even in dimensions $d<3$, potentially driven by superdiffusive classical motion~\cite{GarciaGarcia2008,GarciaGarcia2005}. All these predictions highlight a remarkable feature of the scaling approach: possible dynamical phases can be inferred from classical transport properties and the spectral dimension, without requiring an explicit mapping to a lattice model. 

To assess the robustness of the above semiclassical analysis, we examine the impact of quantum corrections. Quantum effects, such as destructive interference, are known to suppress or even halt transport, raising the question of whether the semiclassical predictions remain valid. For clarity, we focus on the case $d=1$ and $d_e=1$, where classical diffusion is governed by a fractional Fokker–Planck equation \cite{Zaslavsky2002}:
\begin{equation}
  \left(\frac{\partial}{\partial t} - D_{\mathrm{cl}} \frac{\partial^{2 / \mu}}{\partial |p|^{2 / \mu}}\right) P(p, t) = \delta(p) \delta(t).
\end{equation}
Here $\partial^{2/\mu}/\partial|p|^{2/\mu}$ denotes the Riesz fractional derivative, whose Fourier representation is given by multiplication by $-|q|^{2/\mu}$, where $q$ is conjugate to $p$. We consider the infinite momentum axis $p\in\mathbb{R}$, with $P(p,t)\to0$ as $|p|\to\infty$. Applying a Fourier transform from $(p,t)$ space to $(q,\omega)$ space yields the classical propagator:
\begin{equation}
K_0(q,\omega) = \frac{\nu}{-i\omega + D_{\rm cl} |q|^{\frac{2}{\mu}}},
\label{eq:K0}
\end{equation}
where $\nu$ is a constant prefactor related to the density of states. The one-loop-corrected propagator $K(q,\omega)$ modify this classical propagator $K_0(q,\omega)$ via the Dyson equation:
\begin{equation}
K^{-1}(q,\omega) = K_0^{-1}(q,\omega) - \Sigma(q,\omega),
\label{eq:Dyson}
\end{equation}
where the self-energy correction $\Sigma$ accounts for interference corrections originating from intermediate momentum modes $k$:
\begin{equation}
\Sigma(q,\omega=0) = C_0 \int \frac{|q+k|^{2/\mu}-|k|^{2/\mu}}{|k|^{2/\mu}}\,dk.
\label{eq:sigma}
\end{equation}
To evaluate this integral, we use the expansion \cite{Mirlin2000}: 
\begin{equation}
\frac{|q+k|^{2/\mu}}{|k|^{2/\mu}}-1\simeq
\begin{cases}
\frac{2}{\mu}\frac{qk}{k^2}+\frac{1}{\mu}\frac{q^2}{k^2}+\frac{2}{\mu}\left(\frac{1}{\mu}-1\right)\left(\frac{qk}{k^2}\right)^2+\cdots, & |q|\ll |k|,\\
\frac{|q|^{2/\mu}}{|k|^{2/\mu}}, & |q|\gg |k|.
\end{cases}
\end{equation}
The quantum diffusion coefficient $D_{\rm qu}$ to this order is obtained as
\begin{equation}
D_{\rm qu}=D_{\rm cl}-\frac{\nu\Sigma(q,\omega=0)}{|q|^{2/\mu}}=
\begin{cases}
D_{\rm cl}-CN^{2/\mu-1}, & 1<2/\mu<2,\\
D_{\rm cl}-C\ln(|q|N), & 2/\mu=1,\\
D_{\rm cl}-C|q|^{1-2/\mu}, & 0<2/\mu<1.
\end{cases}
\end{equation}
where $C$ is a different constant for each case, $q$ is the external Fourier momentum conjugate to $p$, and the finite momentum-space size $N$ introduces an infrared cutoff $|k|\sim 1/N$ in the loop integral. The quantum correction to the diffusion constant depends critically on the exponent $2/\mu$, which characterizes the scaling behavior of the classical propagator in momentum space. This parameter governs the infrared divergence of the one-loop interference integral: when $2/\mu>1$, the correction grows with system size, indicating strong quantum interference and potential localization. For $2/\mu<1$, the correction is subleading and the classical diffusion remains stable. The marginal case $2/\mu=1$ corresponds to a metal--insulator transition, where the correction becomes logarithmic, superficially resembling that of two-dimensional disordered systems. However, unlike the conventional two-dimensional case, the logarithmic correction does not grow with system size in the long-wavelength limit $|q|\sim 1/N$, where $\ln(|q|N)$ remains of order unity. Thus, $2/\mu$ serves as a key indicator of whether quantum interference destabilizes classical transport.

The one-parameter scaling theory has been tested across various versions of the kicked rotor model. Of particular interest is the kicked rotor with a singular potential \cite{GarciaGarcia2005}, described by the Hamiltonian
\begin{equation}
  H = \frac{\hat{I}^2}{2} + {V}(\hat{\theta})\sum_{n\in\mathbb{Z}}\delta(t-n),
\end{equation}
where the kicking potential $V(\hat{\theta})$ takes the form
\begin{equation}
{V}(\hat{\theta}) = k|\hat{\theta}|^\alpha, \alpha\in\lbrack -1,1 \rbrack, \qquad \text{or} \qquad {V}(\hat{\theta})  = k\log(|\hat{\theta}|).
\end{equation}
Note that the logarithmic singularity corresponds to the limiting case $\alpha=0$ because $\lim_{\alpha\to 0} \frac{|\theta|^\alpha-1}{\alpha} = \log|\theta| $. The classical dynamics of this system exhibits markedly different behavior depending on the singularity exponent $\alpha$.  
For $\alpha>1/2$ the second moment of the momentum increments is finite, $\langle ( \Delta I)^2\rangle < \infty$, leading to normal diffusion $\langle (\Delta I)^2\rangle \sim Dt$, and the corresponding quantum dynamics display Poissonian spectral statistics, consistent with the prediction of the one-parameter scaling theory for localized regimes.  In contrast, for $\alpha < 1/2$ the variance diverges, $\langle (\Delta I)^2\rangle = \infty$, the central limit theorem fails, and the dynamics exhibits superdiffusive (Lévy-type) momentum spreading. According to Eq.~\eqref{eq:scal_fun}, the classical scaling function scales as
\begin{equation}
  \beta_{\rm cl}(g) =  - \frac{\alpha}{1-\alpha}.
\end{equation}
For $\alpha>0$, the scaling theory predicts an insulating phase regardless of quantum corrections. For $\alpha<0$, it predicts a metallic phase, as quantum interference does not qualitatively alter the classical transport. The marginal case $\alpha = 0$ (logarithmic singularity) marks a critical phase. Taken together, these results demonstrate that dynamical localization can be overcome, even in one dimension.

While the one-parameter scaling theory captures how localization and spectral properties depend on the momentum-space size $N$, it is intrinsically a static framework and cannot describe the time evolution of wave packets.  In contrast, recent works extended the scaling approach to explicitly include dynamics~\cite{Chen2023}. They demonstrated that for the logarithmic kicked rotor, where the system resides at the critical point predicted by one-parameter scaling, the quantum dynamics exhibit dynamical finite-size scaling involving both the system size $N$ and the evolution time $t$. This formulation successfully characterizes the temporal spreading and multifractal correlations of critical wave packets, providing a dynamical generalization of the conventional one-parameter scaling framework. This was further generalized to effectively infinite-dimensional Anderson transitions \cite{Chen2024,Chen2024a}, revealing a novel logarithmic multifractal phase. In this regime, critical observables scale with $\ln(N)$ and $\ln(t)$ rather than conventional power laws, marking a distinct universality class of criticality in disordered quantum systems. In short, these advances establish the kicked rotor as a paradigmatic and experimentally accessible platform for exploring Anderson transitions, critical phases and the dynamical facets of quantum criticality.

\subsection{Spin-1/2 kicked rotor}

In previous subsections, we have shown that the on-resonance spin-1/2 kicked rotor can be harnessed to engineer a wide variety of topological phases. Surprisingly, even away from resonance, where the classical system becomes fully chaotic and no Bloch-type quantum numbers exist, topological physics re-emerges in a completely different form. In the nonresonant, chaotic regime, a phenomenon known as Planck's quantum-driven integer quantum Hall effect \cite{Chen2014,Tian2016} closely mimics the integer quantum Hall effect (IQHE) despite the absence of any magnetic field or Fermi statistics. In the following, we first present the conceptual framework that illustrates how topology can emerge from chaos, and then discuss some recent advances.

The quantum dynamics of the model under consideration are governed by
\begin{equation}
i \hbar_{\rm eff}\,\partial_t \tilde{\psi}_t=\hat{H}(t)\,\tilde{\psi}_t,\qquad \hat{H}(t)= \frac{1}{2}\hat{I}_1^2+\sum_\alpha V_\alpha(\hat\theta_1,\hat\theta_2+\tilde{\omega}t)\hat{\sigma}_\alpha
  \sum_{n\in\mathbb{Z}}\delta(t-n),
\label{eq:H-1D}
\end{equation}
where the quantity $\tilde{\omega}/2\pi$ is assumed to be irrational. The spin-dependent kicking potentials $V_\alpha$ exhibit specific parity properties: $V_x$ is odd in $\theta_1$ and even in $\theta_2$; $V_y$ is even in $\theta_1$ and odd in $\theta_2$; and $V_z$ is even in both $\theta_1$ and $\theta_2$.  In the following we focus on the rotor's dimensionless kinetic energy 
\begin{equation}
  E(t)\equiv -\frac{1}{2}\left\langle\langle\psi_t|\partial_{\theta_1}^2|\psi_t\rangle\right\rangle_{\theta_2},
  \label{eq:kineenergy}
\end{equation}
where $\langle\rangle_{\theta_2}$ denotes the average over the prescribed phase $\theta_2$. For simplicity, we assume that the initial state is uniform in $\theta_1$. The long-time behavior of $E(t)$ characterizes the rotor’s transport properties: a vanishing asymptotic growth rate, $\lim\limits_{t\to\infty} \frac{E(t)}{t}=0$, corresponds to bounded motion in angular momentum space, analogous to insulating behavior in disordered electronic systems. Conversely, a nonzero growth rate indicates unbounded motion, mimicking metallic transport. In the following we will see that only for a discrete set of critical values of $\hbar_{\rm eff}$ can $E(t)$ increase linearly at large $t$, which bears a close resemblance to the integer quantum Hall effect.

To see how topological structure can emerge from this one-dimensional model, we first reinterpret the system as a two-dimensional periodic one by treating the parameter $\theta_2$ as a ``virtual" dynamical variable. This construction effectively maps the original one-dimensional system into a higher-dimensional parameter space, allowing us to explore topological features that are otherwise inaccessible in strictly one-dimensional settings. This is achieved by performing the following gauge-like transformation:
\begin{equation}
\hat{H}\;\longrightarrow\;
e^{\tilde{\omega}t\partial_{\theta_2}}\hat{H}e^{-\tilde{\omega}t\partial_{\theta_2}},
\qquad
\tilde{\psi}_t\;\longrightarrow\;
e^{-\tilde{\omega}t\partial_{\theta_2}}\tilde{\psi}_t
\equiv\psi_t.
\label{eq:gauge-transform}
\end{equation}
Substituting Eq.~\eqref{eq:gauge-transform} into Eq.~\eqref{eq:H-1D}, the transformed Schr\"odinger equation becomes
\begin{equation}
i \hbar_{\rm eff}\,\partial_t \psi_t=\left(\frac{1}{2}\hat{I}_1^2+\tilde{\omega}\hat{I}_2 + \sum_\alpha V_\alpha(\hat\theta_1,\hat\theta_2)\hat{\sigma}_\alpha
  \sum_{n\in\mathbb{Z}}\delta(t-n)\right)\psi_t,
\label{eq:H-2D}
\end{equation}
where $\hat{I}_2=-i\hbar_{\rm eff}\partial_{\theta_2}$ is the conjugate variable of $\theta_2$. The Floquet operator for the two-dimensional system reads
\begin{equation}
\hat{U} = \exp(-\frac{i}{\hbar_{\rm eff}}\left(\frac{1}{2}\hat{I}_1^2+\tilde{\omega}\hat{I}_2\right))\exp(-\frac{i}{\hbar_{\rm eff}}\sum_\alpha V_\alpha(\hat{\theta}_1,\hat{\theta}_2)\hat{\sigma}_\alpha).
\label{eq:FSH2D}
\end{equation}
Here, the term $V_x\hat{\sigma}_x+V_y\hat{\sigma}_y$ mimics a 2D ``spin-orbit coupled" electron, with $\Theta\equiv(\theta_1,\theta_2)$ playing the role of its ``momentum". Meanwhile, $V_z\sigma_z$ acts as a ``Zeeman coupling", explicitly breaking the effective time-reversal symmetry $i\hat{\sigma}_y\mathcal{K}$, where $\mathcal{K}$ denotes complex conjugation in the angular-momentum basis. The full antiunitary operation acts as $\hat\theta_{1,2}\to-\hat\theta_{1,2}$ and $\hat I_{1,2}\to\hat I_{1,2}$. The first exponential in Eq.~\eqref{eq:FSH2D} is diagonal in the angular-momentum basis $|n_1,n_2\rangle$, producing momentum-dependent phases $\exp[-i(\hbar_{\rm eff}n_1^2/2+\tilde{\omega}n_2)]$. When $\hbar_{\rm eff}$ and $\tilde{\omega}$ are incommensurate with $2\pi$, these phases form a nonperiodic pattern in the momentum lattice $N\equiv(n_1,n_2)$, effectively mimicking disorder, where $n_1$ and $n_2$ are the eigenvalues of $\hat{n}_1=\hat{I}_1/\hbar_{\rm eff}$ and $\hat{n}_2=\hat{I}_2/\hbar_{\rm eff}$, respectively. Thus, this analogy already hints at an IQHE-like topological behavior driven purely by quantum interference. 

\begin{figure}
\centering
\includegraphics[width=0.40\textwidth]{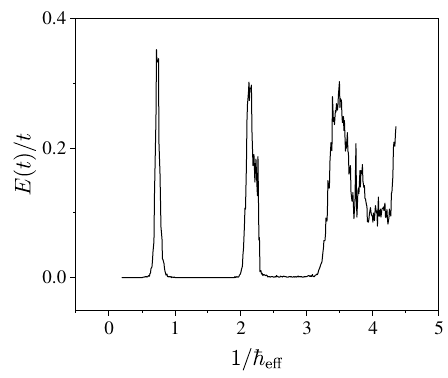}
\caption{Energy growth rate of the spin-$1/2$ kicked rotor for different values of $\hbar_{\rm eff}$. 
The kicking potentials are set as 
$V_x = \tfrac{2\arctan(2|\bm{d}|)}{|\bm{d}|}\,d_1$, 
$V_y = \tfrac{2\arctan(2|\bm{d}|)}{|\bm{d}|}\,d_2$, and 
$V_z = \tfrac{2\arctan(2|\bm{d}|)}{|\bm{d}|}\,d_3$, 
with $\bm{d} = (d_1,d_2,d_3) = \bigl(\sin\theta_1,\;\sin\theta_2,\;0.8\,(1-\cos\theta_1-\cos\theta_2)\bigr)$. The pronounced peaks in $E(t)/t$ correspond to critical states separating topologically distinct insulating phases, analogous to the longitudinal conductivity peaks at plateau transitions in the IQHE.
(Adapted from \cite{Tian2016}.)}
\label{fig:spinhalf}
\end{figure}

To investigate the long-time transport properties of the two-dimensional system, we express the kinetic energy in terms of the retarded--advanced Floquet correlation functions introduced earlier. In the angular-momentum basis $|N,s\rangle$, with $N=(n_1,n_2)$ and $s=\uparrow,\downarrow$, the corresponding density correlation function is
\begin{equation}
K_\omega(Ns_+s_-,N's'_+s'_-) = \left\langle\langle N,s_+|G^+(\omega_+)|N',s'_+\rangle\langle N',s'_-|G^-(\omega_-)|N,s_-\rangle\right\rangle_{\omega_0},
\label{eq:Komega}
\end{equation}
where $\omega_\pm=\omega_0\pm\omega/2$, and $\langle\cdots\rangle_{\omega_0}$ denotes the quasienergy average. For the initial state $|\psi_0\rangle=\sum_s\psi_{0,s}|N=0,s\rangle$, the kinetic energy in Eq.~\eqref{eq:kineenergy} becomes
\begin{equation}
E(t)=\frac12\int_0^{2\pi}\frac{d\omega}{2\pi}e^{-i\omega t}\sum_N n_1^2\sum_{s,s'_+,s'_-}K_\omega(Nss,0s'_+s'_-) \psi_{0,s'_+}\psi_{0,s'_-}^*.
\label{eq:two_p_GF}
\end{equation}
Thus, the long-time energy growth is determined by the correlations between forward and backward Floquet propagation in the effective two-dimensional angular-momentum space.

Following the field-theoretical construction outlined previously, the quasienergy-averaged correlation function in Eq.~\eqref{eq:Komega} can be expressed as a functional integral over supermatrix fields $Z$ and $\widetilde Z$,
\begin{equation}
K_\omega\sim\int\mathcal D(Z,\widetilde Z)e^{-S[Z,\widetilde Z]}(\cdots),
\label{eq:fieldint}
\end{equation}
where the supermatrix fields describe correlations between retarded and advanced propagation amplitudes. In the present model, these fields additionally carry spin indices and are defined on the two-dimensional angular-momentum lattice. The connection to an effective nonlinear sigma model follows by identifying the slowly varying fields that govern the long-time dynamics. In the strongly chaotic regime, correlations associated with different angular-momentum states decay rapidly, allowing the off-diagonal components $Z_{N,N'}$ with $N\neq N'$ to be integrated out. Furthermore, decomposing the remaining fields in spin space as $Z=Z_0\sigma_0+\bm{Z}\cdot\bm{\sigma}$, one finds that the spin-dependent components $\bm{Z}$ are massive. The low-energy theory is therefore governed by a smooth, spin-singlet field $Z(N)$, reflecting the conservation of probability even though spin polarization is not conserved.

As in the conventional kicked rotor, the nonlinear structure of the soft modes can be parametrized by a supermatrix $Q(N)=T^{-1}(N)\Lambda T(N)$ satisfying $Q^2=1$, where $\Lambda$ distinguishes the retarded and advanced sectors. However, the present system differs from the conventional rotor in two important respects: its effective angular-momentum space is two-dimensional, and the spin-dependent kicking breaks time-reversal symmetry. Consequently, the long-wavelength expansion gives a nonlinear sigma model of the unitary symmetry class, with an additional topological term~\cite{Chen2014,Tian2016},
\begin{equation}
S[Q]=\frac14\int d^2N\,\operatorname{str}\left[-\sigma(\nabla Q)^2+\sigma_{\mathrm H}Q(\partial_{n_1}Q)(\partial_{n_2}Q)-2i\omega Q\Lambda\right],
\label{eq:spin_sigma}
\end{equation}
where $\operatorname{str}$ denotes the supertrace over the internal indices, $\sigma$ is the bare diffusion coefficient, and $\sigma_{\mathrm H}$ determines the strength of the topological term. These two parameters play roles analogous to the longitudinal and Hall conductivities in the conventional integer quantum Hall effect. In particular, $2\pi\sigma_{\mathrm H}$ acts as the topological theta angle.

The origin of this topological term can be understood from the geometry of the $Q$ field. The bosonic part of the target manifold is equivalent to $H^2\times S^2$, where $H^2$ is a noncompact hyperbolic space and $S^2$ is a compact sphere. These spaces arise from the bosonic and fermionic sectors of the supersymmetric theory, rather than directly from the physical spin degree of freedom. Since $Q(N)$ is defined on a two-dimensional angular-momentum space, field configurations approaching a constant at infinity define mappings from the compactified $N$ space to this target manifold. Such mappings are classified by
\begin{equation}
\pi_2(H^2\times S^2)=\pi_2(H^2)\times\pi_2(S^2)=\pi_2(S^2)=\mathbb Z.
\label{eq:homotopy1}
\end{equation}
Here $H^2$ is contractible and does not contribute to the nontrivial topology. The integer-valued winding number originates from the compact $S^2$ sector and allows the topological term in Eq.~\eqref{eq:spin_sigma} to distinguish different classes of field configurations.

The resulting transport behavior is governed by the renormalization of the two couplings $\sigma$ and $\sigma_{\mathrm H}$. The former controls the diffusive energy growth at short times, while the latter determines the topological response. In the strongly chaotic regime and for sufficiently small $\hbar_{\rm eff}$, their bare values obey the characteristic scalings
\begin{equation}
\sigma\sim\hbar_{\rm eff}^{-2},\qquad \sigma_{\mathrm H}\sim\hbar_{\rm eff}^{-1}.
\label{eq:spin_sigma_scaling}
\end{equation}
At long times, quantum interference and nonperturbative topological fluctuations renormalize these parameters. The resulting two-parameter renormalization-group flow exhibits insulating fixed points with vanishing longitudinal transport and integer-valued topological index $\sigma_{\mathrm H}^*$. Between neighboring insulating phases, critical points support a finite, universal asymptotic energy growth rate,
\begin{equation}
E(t)\sim\sigma^*t,\qquad \sigma^*>0.
\label{eq:spin_critical}
\end{equation}
As $\hbar_{\rm eff}$ is varied, successive transitions occur between insulating phases characterized by different integer values of $\sigma_{\mathrm H}^*$. This produces the Planck's quantum-driven integer quantum Hall effect illustrated in Fig.~\ref{fig:spinhalf}. The correspondence between its transport characteristics and those of the conventional integer quantum Hall effect is summarized in Tab.\ref{tab:spinrotor}. 

\begin{table}[t]
\centering
\caption{The analogy between Planck's quantum-driven integer quantum Hall effect and conventional integer Quantum Hall effect. Here, $\sigma_H^*$ denotes the integer-valued topological quantum number characterizing the dynamically localized phases, whereas $\sigma^*$ denotes the universal asymptotic energy growth rate at the critical metallic points. (Adapted from \cite{Tian2016}.)}
\begin{tabular}{lcc}
\hline\hline
 & Planck's quantum-driven IQHE & Conventional IQHE \\
\hline
System & Spin-$\tfrac{1}{2}$ quasiperiodic quantum kicked rotor & 2D electron gas \\
Driving parameter & Planck’s quantum $\hbar_{\rm eff}$ & Magnetic field  \\
Characteristic of dissipation & Energy growth rate $\lim_{t\to\infty}\dfrac{E(t)}{t}$ & Longitudinal conductivity \\
Characteristic of topology & Hidden quantum number $\sigma_{\mathrm{H}}^{*}$ & Quantized Hall conductivity \\
Characteristic of insulator & $\lim_{t\to\infty}\dfrac{E(t)}{t}=0$ & Vanishing longitudinal conductivity \\
Characteristic of metal & $\lim_{t\to\infty}\dfrac{E(t)}{t}=\sigma^{*}$ & Finite longitudinal conductivity \\
\hline\hline
\end{tabular}
\label{tab:spinrotor}
\end{table}

The above topological structure relies on the existence of effective two-dimensional chaotic dynamics, which requires the modulation frequency $\tilde{\omega}/(2\pi)$ to be irrational. When this ratio becomes rational, the Floquet operator acquires translational symmetry along the $n_2$ direction. The corresponding Bloch decomposition reduces the dynamics to a family of quasi-one-dimensional subsystems, rather than a genuinely two-dimensional diffusive system. Consequently, the effective field theory no longer supports the two-dimensional topological term in Eq.~\eqref{eq:spin_sigma}. This can also be understood from the trivial first homotopy group of the target manifold,
\begin{equation}
\pi_1(H^2\times S^2)=\pi_1(H^2)\times\pi_1(S^2)=0.
\label{eq:homotopy2}
\end{equation}
The absence of the topological term reduces the long-wavelength theory to a quasi-one-dimensional localization problem, and the Planck's quantum-driven integer quantum Hall effect disappears. In particular, the critical linear energy growth is replaced by a crossover toward dynamical localization in the $n_1$ direction.

The above analysis relies crucially on the assumption of strong chaos, which ensures rapid decay of correlations and justifies the reduction to a smooth field $Z(N)$. Interestingly, recent studies have revealed that phenomena resembling the integer quantum Hall effect can also emerge in nonchaotic systems~\cite{Guarneri2020}. Specifically, the spin-Maryland model, an integrable, self-dual extension of the standard Maryland rotor~\cite{Grempel1982}, exhibits quantized transitions between dynamically localized and delocalized phases~\cite{Guarneri2020}. There, the quantization originates not from ergodic mixing, but from an exact spectral self-duality that maps the angular and momentum representations onto each other, leading to a topological classification of dynamical phases, where each phase is indexed by a topological invariant in the parameter (effective Planck constant) space. These findings demonstrate that topological quantization can emerge through distinct physical routes: either from chaotic mixing in the spin-$1/2$ kicked rotor, or from exact self-duality in the integrable spin-Maryland model. Together, they highlight the remarkable richness of dynamical behavior in spinful kicked-rotor systems, where chaos, integrability, and topology are deeply intertwined.

\subsection{Coupled kicked rotors}
In the above discussion, our attention has primarily been restricted to the (effective) single-body kicked-rotor model, which already exhibits a variety of nontrivial dynamical behaviors. Recent studies, however, have extended this framework to the many-body domain, giving rise to the coupled kicked-rotor system. Even when only two rotors are coupled, the dynamical landscape becomes considerably richer: the interplay between periodic driving and inter-rotor interaction generates complex entanglement dynamics, which plays a central role in diagnosing thermalization and localization in interacting quantum systems~\cite{Abanin2019}.  In what follows, we first introduce the transport and entanglement dynamics in the coupled kicked rotor through a representative example~\cite{Paul2020}, and then briefly outline some of the recent frontiers in coupled-kicked-rotor model. 

The Floquet operator for the two-body coupled kicked rotor we consider here is given by 
\begin{equation}
\hat{U} = \exp\left(-i\left(\frac{\hat{I}_1^2}{2} + \frac{\hat{I}_2^2}{2}\right)\right)\exp\left(-i\left(k_1\cos\hat\theta_1 + k_2\cos\hat\theta_2 +\xi\cos\left(\hat\theta_1-\hat\theta_2\right)\right)\right) ,
\label{eq:Cou_Rot}
\end{equation}
where we have set $\hbar=1$ throughout this subsection. Hereafter we focus on the parameter regime in which the single-body kicking strengths $k_j$ are sufficiently large, while the inter-rotor coupling strength $\xi$ is small. We take the initial state to be a product states of the form 
\begin{equation}
|\Psi(0)\rangle = |\psi_1(0)\rangle\otimes |\psi_2(0)\rangle , 
\end{equation}
where $ |\psi_j(0)\rangle$ is a coherent state of the $j$th rotor.

We begin by examining the energy-transport properties of the $j$th rotor, defined by $E_j(t) =\langle \Psi(t)|\frac{I_j^2}{2}|\Psi(t)\rangle$. When $\xi=0$, each rotor exhibits dynamical localization independently at around its diffusion time scale $t_b$. As discussed previously, the destructive quantum interference between multiple scattering paths in angular momentum space leads each rotor to dynamical localization, thereby suppressing energy diffusion. However, the coupled system (\ref{eq:Cou_Rot}) modifies this picture in a subtle way. The interaction 
\begin{equation}
\hat{H}_{I} = \xi\cos(\hat\theta_1-\hat\theta_2)\sum_{n\in\mathbb{Z}} \delta(t-n)
\end{equation}
couples the two rotors in angular position space at each kick. From a semiclassical viewpoint, one can regard rotor~2 as a time-dependent ``force'' acting on rotor~1 through the phase factor $e^{-i\xi\cos(\theta_1-\theta_2)}$, and vice versa. For fully chaotic single-rotor dynamics, the relative phase $\theta_1-\theta_2$ fluctuates irregularly, motivating an effective phase-noise picture for the reduced dynamics of each rotor. Such effective phase modulation can weaken the quantum interference responsible for dynamical localization. However, the joint two-rotor system remains fully coherent under unitary evolution, and this picture does not by itself establish delocalization of the full system.

This destabilization mechanism can be further clarified through the lattice-model mapping perspective. A single quantum kicked rotor maps onto a one-dimensional tight-binding Anderson model with static disorder in the on-site terms, where dynamical localization corresponds to Anderson localization in angular momentum space. In contrast, the coupled kicked rotors yield a Floquet matrix that maps onto an effective two-dimensional lattice model~\cite{Borgonovi1995}. Importantly, the present coupling term $\xi\cos(\hat\theta_1-\hat\theta_2)$ introduces additional off-diagonal hopping between different momentum states, rather than simply renormalizing the on-site disorder. Such additional hopping channels modify the transport properties of the effective two-dimensional lattice. However, the nontrivial correlations in this lattice require care when applying conventional localization arguments, and enhanced transport over finite times does not necessarily imply asymptotic delocalization.

The above scenario naturally introduces two distinct and physically different time scales. The first one is the single-rotor quantum break time $t_b$, which is the time scale on which coherent backscattering in momentum space builds up and the classical diffusion is suppressed for an isolated kicked rotor. The second time scale, denoted by $t^\ast(\xi)$, is a coupling-induced crossover time. It characterizes the end of the intermediate dynamical localization regime and the onset of renewed energy growth observed in the numerical dynamics. Unlike $t_b$, the crossover time $t^\ast$ increases as $\xi$ decreases and is expected to diverge in the limit $\xi\to0$. For weak coupling one has the hierarchy $t_b \ll t^\ast(\xi)$.

\begin{figure}
\centering
\includegraphics[width=0.40\textwidth]{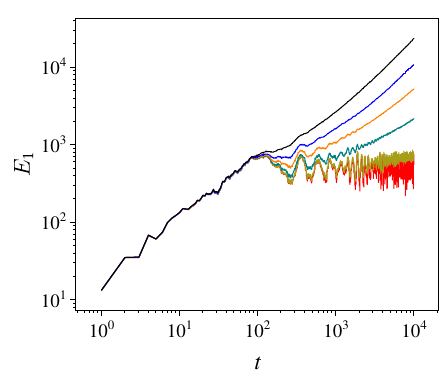}
\caption{Average energy of the first rotor, with $k_1=9.0$ and $k_2=10.0$. 
The coupling strengths are, from bottom to top, $\xi=0,\,0.01,\,0.03,\,0.05,\,0.07,$ and $0.1$.
( Adapted from \cite{Paul2020}).}
\label{fig:couple_energy}
\end{figure}

Therefore, these considerations lead to a three-stage scenario for the energy growth of each rotor under weak but finite coupling: (1) For short times $t\lesssim t_b$, the quantum dynamics closely follows the classical chaotic diffusion, $E_j(t)\simeq D_{\mathrm{cl}}t$, where $D_{\mathrm{cl}}$ denotes the classical diffusion coefficient. (2) At intermediate times $t_b\lesssim t\lesssim t^\ast(\xi)$, a regime of dynamical localization appears. Quantum interference still partially suppresses transport, so that $E_j(t)$ remains approximately constant over a broad time window. (3) For $t\gtrsim t^\ast(\xi)$, the coupling-induced phase modulation is associated with renewed energy growth, which exhibits an approximately diffusive behavior, $E_j(t)\simeq D_{\rm q}t$, over the numerically accessible time scales, with $D_{\rm q}$ denoting the effective quantum diffusion coefficient. This three-stage scenario is supported by the numerical results shown in Fig.~\ref{fig:couple_energy}. Nevertheless, the observed finite-time diffusion does not exclude the possibility of eventual localization on much longer time scales.

We now turn to the entanglement dynamics. When $\xi = 0$, the system consists of two uncoupled kicked rotors and therefore no entanglement is generated during the time evolution. Introducing a nonzero coupling $\xi$ induces correlations between the two subsystems, leading to the development of entanglement. 
For pure global states of a bipartite system, the standard measure of entanglement is the von Neumann entropy of the reduced density matrix of either subsystem, often referred to as the entanglement entropy~\cite{Plenio2007}, defined as
\begin{equation}
  S_\mathrm{vN}(t)=- \operatorname{Tr}\left[\hat\rho_1(t) \log \hat\rho_1(t)\right], 
  \label{eq:SvN-def}
\end{equation}
where $\hat\rho_1(t) = \operatorname{Tr}_2  |\Psi(t)\rangle\langle\Psi(t)|$ denotes the reduced density matrix of rotor~1 obtained by tracing out the degrees of freedom of rotor~2. For the bipartite pure state considered here, the von Neumann entropy provides a complete characterization of bipartite entanglement, capturing both its growth and eventual saturation. For analytic purposes, however, it is often convenient to employ the linear entropy,
\begin{equation}
  S_\mathrm{lin}(t)=1 - \operatorname{Tr}\left[ \hat\rho_1^2(t)\right].
  \label{eq:Slin-def}
\end{equation}
Although $S_{\mathrm{lin}}$ may not share the same scaling form as the von Neumann entropy, it can faithfully capture the dynamical crossover between different entanglement-growth regimes and allows for perturbative treatments in the weak-coupling regime.

\begin{figure}
\centering
\includegraphics[width=0.55\textwidth]{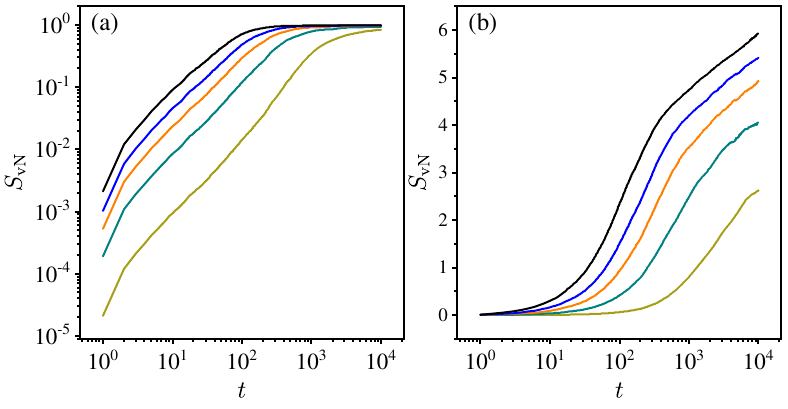}
\caption{Time evolution of the von Neumann entanglement entropy. 
Panel (a) shows the results on a log-log scale, while panel (b) uses a log-linear scale. 
Parameters are $k_1=9.0$ and $k_2=10.0$. 
The coupling strengths are, from bottom to top, $\xi=0.01,\,0.03,\,0.05,\,0.07,$ and $0.1$. 
(Adapted from \cite{Paul2020}.)}
\label{fig:couple_ent}
\end{figure}

Since the generation of entanglement is solely due to the inter-rotor coupling, it is convenient to analyze the dynamics in the interaction picture with respect to the uncoupled Hamiltonian. The interaction-picture transformation is defined as 
\begin{equation}
\hat\rho^I(t)=\left(\left(\exp\left(-i\left(\frac{\hat I_1^2}{2} + \frac{\hat I_2^2}{2}\right)\right)\exp\left(-i\left(k_1\cos\hat\theta_1 + k_2\cos\hat\theta_2 \right)\right)\right)^t\right)^\dagger \hat\rho(t) \left(\exp\left(-i\left(\frac{\hat I_1^2}{2} + \frac{\hat I_2^2}{2}\right)\right)\exp\left(-i\left(k_1\cos\hat\theta_1 + k_2\cos\hat\theta_2 \right)\right)\right)^t,
\end{equation}
and 
the total density matrix obeys
\begin{equation}
\frac{d\hat\rho^I(t)}{dt} = -i \lbrack \hat H_I(t),\hat\rho^I(t)]\rbrack.
\end{equation}
As the initial state is a product state, we have $\hat\rho^I(0) = |\psi_1(0)\rangle\langle\psi_1(0) |\otimes |\psi_2(0)\rangle\langle \psi_2(0)|$. Performing formal integration and iteration and expanding to second order in $\xi$ gives the result
\begin{equation}
\hat\rho^I(t) = \hat\rho^I(0) -i\xi\sum_{r=1}^{t}[\hat F(r),\hat\rho^I(0)] + \left(i\xi\right)^2 \sum_{r=1}^{t}\sum_{s=1}^{r-1} [\hat F(s),[\hat F(r),\hat\rho^I(0)]],
\label{eq:rho-2nd}
\end{equation}
where $\hat F(r)=\cos\left(\hat\theta_1(r)-\hat\theta_2(r)\right)$. Tracing over the second rotor and using  the definition of linear entropy, one obtains
\begin{equation}
S_\mathrm{lin}(t) = \xi^2  \sum_{r,s=1}^{t} C(r,s),
\label{eq:Slin-Ct}
\end{equation}
where $C(r,s)$ is a two-time correlation function of the interaction operator, defined as
\begin{equation}
C(r,s) = \operatorname{Tr}_1 \!
\left[ \operatorname{Tr}_2 \Big( \big[\hat F(r),\hat\rho^I(0)\big] \Big)\; \operatorname{Tr}_2\Big(\big[\hat F(s),\hat\rho^I(0)\big]\Big)\right].
\label{eq:Crs-def}
\end{equation}
For fully chaotic single-rotor dynamics these correlations decay rapidly with $|r-s|$, and one finds numerically that $C(t)\propto t$. Consequently, the early-time entanglement growth is linear \footnote{It should be noted that Ref.~\cite{Gong2003} proved that, on short time scales, the entanglement entropy cannot exhibit linear growth. In the present model, however, the characteristic time scale is extremely short, and the time window considered here lies beyond this transient regime.},
\begin{equation}
S_\mathrm{lin}(t)\propto \xi^2 t,
\end{equation}
with a slope determined by the interaction strength $\xi$. This linear growth in the short-time regime is confirmed by the numerical results shown in fig.~\ref{fig:couple_ent}(a).

The above perturbative expression is derived under the assumption that the total state remains close to a product state, i.e. that the correction 
$\delta\hat \rho_1^I(t)$ remains small compared to the unperturbed reduced density matrix $\hat \rho_1^{(0)}(t)$. This assumption breaks down once strong entanglement is formed at long times. In the diffusive regime of energy transport discussed above, the reduced phase-space distribution of each rotor becomes approximately Gaussian in angular momentum and nearly uniform in the angle coordinate. In this regime the reduced state is well described by a smooth Husimi function $\mathcal{H}(\theta_1,I_1)$, and the linear entropy can be expressed as \cite{Nag2001}
\begin{equation}
S_\mathrm{lin} = 1-\int\frac{d\theta_1\,dI_1}{2\pi}\mathcal{H}^2(\theta_1,I_1).
\end{equation}
Approximating the Husimi function in the diffusive regime by a Gaussian in angular momentum, $\mathcal{H}(\theta_1,I_1)\simeq\sqrt{\frac{1}{2\pi D_{\rm q} t}}\exp\!\left(-\frac{I_1^2}{2D_q t}\right)$, where $D_{\rm q}$ denotes the effective quantum diffusion coefficient in the observed transport regime, one obtains
\begin{equation}
S_\mathrm{lin}(t)= 1-\left(\frac{1}{4\pi D_{\rm q} t}\right)^{1/2}.
\label{eq:Slin-asymp}
\end{equation}
Thus, in contrast to the early-time linear growth, the linear entropy approaches its maximal value only algebraically at long times, reflecting the diffusive broadening of the reduced phase-space distribution. It should be noted that the von Neumann entropy $S_{\rm vN}$ behaves differently from $S_{\rm lin}$ in this regime, since $S_{\rm lin}$ is bounded by unity by definition, whereas $S_{\rm vN}$ has no finite upper bound in the infinite-dimensional Hilbert space of the rotor considered here. In the diffusive regime, the effective number of participating Schmidt modes, characterized by the inverse purity of the reduced density matrix \cite{qcbook}, can be estimated from the Husimi distribution as
$N_{\rm eff}(t)\simeq \left( \int\frac{d\theta_1\,dI_1}{2\pi}\,\mathcal{H}^2(\theta_1,I_1) \right)^{-1}
\sim \sqrt{D_{\rm q}t}$.
Since the von Neumann entropy scales as $S_{\rm vN}\sim \ln N_{\rm eff}$, one obtains the logarithmic growth in the diffusive regime,
\begin{equation}
S_{\rm vN}(t)\simeq \frac{1}{2}\ln t + \mathrm{const},
\end{equation}
which is also confirmed in Fig.\ref{fig:couple_ent}(b).

In the above, we have focused on a specific model of coupled kicked rotors. More generally, coupled rotor systems provide a flexible and unifying framework for exploring interaction effects in periodically driven quantum systems. In particular, many-body coupled kicked rotors open the door to addressing a range of outstanding questions at the intersection of chaos, transport, and nonequilibrium dynamics. Within this class of models, interaction effects may lead to rich dynamical phenomena, including interaction-induced delocalization \cite{Rylands2020,SeeToh2022,Cao2022}, subdiffusive transport \cite{Russomanno2021}, and the emergence of Arnold diffusion in high-dimensional phase-space \cite{Schmidt2023}. Overall, coupled kicked rotors offer a minimal and powerful setting to uncover universal features of interaction-driven dynamics in quantum systems.

\subsection{Non-Hermitian kicked rotor}

In recent years, the rapid progress of non-Hermitian physics has offered new perspectives for exploring exotic quantum dynamics beyond the Hermitian paradigm. Within this framework, non-Hermitian extensions of the kicked rotor provide a minimal setting to investigate how quantum-interference phenomena, such as dynamical localization and quantum resonances, are modified by gain--loss mechanisms. We use a $\mathcal{PT}$-symmetric kicked rotor as an example to briefly introduce the interplay between the kicked rotor and non-Hermitian physics~\cite{Longhi2017}, followed by a short outlook.

A representative example of a non-Hermitian kicked rotor is the $\mathcal{PT}$-symmetric kicked rotor described by the Floquet operator
\begin{equation}
\hat{U}=\exp\!\left(-i\frac{\hat{I}^2}{2}\,T\right)\exp\!\left[-ik\left(\cos\hat\theta+i\gamma\sin\hat\theta\right)\right],
\label{eq:PTKR-U}
\end{equation}
where $k$ denotes the Hermitian kicking strength and $\gamma$ controls the amplitude of balanced gain and loss. The real and imaginary parts of the kicking potential are even and odd in $\theta$, respectively, ensuring $\mathcal{PT}$ symmetry of the periodically driven Hamiltonian.

A central observation for this model is that, when $|\gamma|<1$, the complex kicking potential can be rewritten as a complex-shifted function. By introducing
\begin{equation}
k_0=k\sqrt{1-\gamma^2}, \qquad \tanh\eta=\gamma,
\label{eq:K0-eta-def}
\end{equation}
the kicking potential can be transformed to
\begin{equation}
k\big(\cos\hat\theta+i\gamma\sin\hat\theta\big)=k_0\cos(\hat\theta-i\eta),
\label{eq:kick-shifted}
\end{equation}
i.e., the non-Hermitian kick is equivalent to an imaginary shift $\theta\mapsto\theta-i\eta$, accompanied by a renormalization of the kicking strength, $k\to k_0$. This immediately suggests a mapping to a Hermitian kicked rotor. Let
\begin{equation}
\hat{U}_0=\exp\!\left(-i\frac{\hat{I}^2}{2}\,T\right)\exp\!\left(-ik_0\cos\hat\theta\right)
\label{eq:U0-Herm}
\end{equation}
be the unitary Floquet operator of the Hermitian kicked rotor with effective kick $k_0$. The imaginary shift can be implemented by a non-unitary similarity transformation generated by $\hat I$. Defining
\begin{equation}
\hat S\equiv\exp(\eta\hat I),
\label{eq:S-def}
\end{equation}
one has $\hat S\hat\theta\hat S^{-1}=\hat\theta-i\eta$ and hence $\hat S f(\hat\theta)\hat S^{-1}=f(\hat\theta-i\eta)$ for any analytic function $f$, while $\hat S\hat I\hat S^{-1}=\hat I$. Since $\hat S$ commutes with the free-rotation factor, it follows that
\begin{equation}
\hat{U}=\hat S\,\hat U_0\,\hat S^{-1},
\label{eq:similarity}
\end{equation}
so that the $\mathcal{PT}$-symmetric Floquet operator is formally similar to a unitary one. At the purely algebraic level, this indicates that the quasienergy spectrum can remain real. For the non-Hermitian case, the dimensionless quasienergy is generally complex, with its real and imaginary parts denoted by $E_r$ and $E_i$, respectively, and the corresponding Floquet eigenvalue given by $\lambda=e^{-i(E_r+iE_i)}$.

\begin{figure}
\centering
\includegraphics[width=0.40\textwidth]{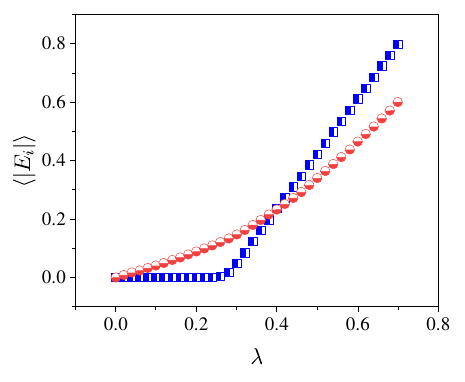}
\caption{Average absolute imaginary part of the quasienergy spectrum for $k=3$. Blue squares correspond to $T=1.4$ (localized regime), while red circles correspond to $T=\pi/3$ (resonant regime).}
\label{fig:NH_spectrum}
\end{figure}

However, the mapping is physically meaningful only if the similarity transformation acts within the Hilbert space of normalizable states. In the angular-momentum basis $|m\rangle$ with $\hat I|m\rangle=m|m\rangle$, one has
\begin{equation}
\hat S|m\rangle=e^{\eta m}|m\rangle,
\end{equation}
so the transformation exponentially reweights momentum components. Consequently, the mapping in Eq.~(\ref{eq:similarity}) is meaningful only if the transformed Floquet eigenstates remain normalizable. This requirement marks the crucial distinction between nonresonant and resonant dynamics.

For the nonresonant regime, where the Hermitian kicked rotor $\hat U_0$ exhibits dynamical localization in angular momentum space, its Floquet eigenstates are exponentially localized: $|\phi_m^{(0)}|\sim\exp(-|m-m_0|/\xi_L)$, with $\xi_L$ the localization length. Under the transformation $|\phi^{(0)}\rangle\to\hat S|\phi^{(0)}\rangle$, the momentum amplitudes acquire a factor $e^{\eta m}$, and remain normalizable provided the localization-induced decay dominates over the exponential reweighting. This yields the condition
\begin{equation}
|\eta|<\xi_L^{-1},
\label{eq:eta-xi}
\end{equation}
which provides an estimate of the finite $\mathcal{PT}$ threshold $\gamma_{\rm PT}$ through $\eta=\operatorname{artanh}\gamma$,
\begin{equation}
\gamma_{\rm PT}\sim\tanh(\xi_L^{-1}).
\label{eq:PT-threshold}
\end{equation}
Consequently, in the dynamically localized regime there exists a transition from an unbroken $\mathcal{PT}$ phase (all quasienergies real) to a broken $\mathcal{PT}$ phase (complex-conjugate quasienergy pairs, defined modulo $2\pi/T$). In this viewpoint, localization stabilizes the real Floquet spectrum by ensuring that the transformed localized eigenstates remain normalizable.

The resonant regime can be understood from the same criterion, Eq.~\eqref{eq:PT-threshold}. However, at generic quantum resonances the Hermitian kicked rotor does not localize; Floquet eigenstates become extended in momentum space and the effective localization length diverges, $\xi_L\to\infty$. In this case the normalizability condition (\ref{eq:eta-xi}) cannot be satisfied for any nonzero $\eta$, so the similarity mapping (\ref{eq:similarity}) breaks down immediately. As a result, the Floquet spectrum becomes complex for arbitrarily small non-Hermiticity, so that the resonant rotor generally supports only the broken-$\mathcal{PT}$ phase, except at special quantum antiresonances. This analysis is corroborated by the numerical results shown in Fig.~\ref{fig:NH_spectrum}.

An interesting dynamical consequence of the broken-$\mathcal{PT}$ resonant dynamics is the emergence of ratchet acceleration in angular momentum space. Taking the principal quantum resonance as a representative example, where the kicking period satisfies $T=4\pi$ and the Floquet operator simplifies to a purely kick-induced evolution,
\begin{equation}
\hat U=\exp\!\left[-ik\left(\cos\hat\theta+i\gamma\sin\hat\theta\right)\right],
\label{eq:resonant-U}
\end{equation}
and expanding the complex kicking potential as
\begin{equation}
\cos\theta+i\gamma\sin\theta=\frac{1}{2}\big[(1+\gamma)e^{i\theta}+(1-\gamma)e^{-i\theta}\big],
\label{eq:harmonics}
\end{equation}
we can observe that non-Hermiticity introduces an intrinsic asymmetry between forward and backward momentum-transfer channels. In contrast to the nonresonant case, where quantum interference suppresses long-range transport, at quantum resonance this asymmetry is not averaged out by localization effects.

\begin{figure}
\centering
\includegraphics[width=0.40\textwidth]{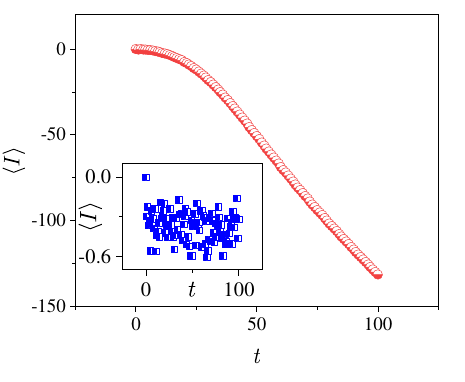}
\caption{Time evolution of the average momentum for $k=3$ and $\gamma=1/30$. Red circles correspond to $T=\pi/3$, and the inset shows the result for $T=1$. (Adapted from \cite{Longhi2017}.)}
\label{fig:NH_ratchet}
\end{figure}

The translational invariance of the resonant Floquet operator permits Bloch eigenstates labelled by a quasi-momentum $q$. The corresponding Floquet eigenvalues define a complex quasienergy dispersion, whose imaginary part leads to mode-dependent amplification. As a result, the long-time dynamics is governed by Bloch components near the wave number $q^\ast$ that maximizes the imaginary part of the quasienergy, i.e., the components with the largest net gain. The real part of the dispersion then directly determines the transport properties of these selected components. Consequently, initially localized states develop a drifting momentum-space wave packet with a nonzero mean angular momentum (as confirmed in Fig.~\ref{fig:NH_ratchet}),
\begin{equation}
\langle I(t)\rangle=\frac{\langle\psi(t)|\hat I|\psi(t)\rangle}{\langle\psi(t)|\psi(t)\rangle}\sim vt,
\end{equation}
where $v=\left.\partial_q E_r(q)\right|_{q=q^\ast}$ is the drift velocity per kick, and the expectation value is normalized by the instantaneous state norm $\langle\psi(t)|\psi(t)\rangle$.

Beyond the interplay between non-Hermiticity and quantum-interference phenomena such as dynamical localization and quantum resonance, non-Hermitian kicked rotors provide a natural and versatile framework for exploring quantum--classical correspondence in driven non-Hermitian systems~\cite{Hall2023}. Moreover, they have been employed to investigate how non-unitarity fundamentally alters the dynamics of out-of-time-ordered correlators~\cite{Zhao2022,Zhao2023}, universal spectral properties~\cite{Li2024}, and topological phases~\cite{Zhou2019}. Taken together, these features establish the non-Hermitian kicked rotor as a minimal and powerful testbed for probing the interplay between chaos, topology, and gain--loss-induced mode selection in driven quantum systems.


\section{Conclusion and Outlook}\label{sec:conclusion}

The kicked rotor stands as one of the most foundational paradigms in the theory of dynamical systems. Its enduring significance lies in the striking contrast between the extreme simplicity of its formulation and the exceptional richness of the dynamics it supports. Originally introduced as a minimal model of classical chaos, the kicked rotor has repeatedly re-emerged over the past decades as a universal reference system for exploring new physical phenomena, including quantum chaos, dynamical localization, Floquet topology, and non-Hermitian physics. Remarkably, this breadth is achieved without invoking disorder, interactions, or complex geometries: a single periodically kicked degree of freedom suffices to generate a hierarchy of dynamical regimes whose structure continues to shape our understanding of complex classical and quantum dynamics. 
It is precisely this combination of formal minimalism and conceptual universality that has established the kicked rotor as a cornerstone of modern nonequilibrium physics.

Given the breadth of the field, several important directions could only be briefly addressed or remain beyond the scope of the present review. For the original kicked rotor, we have not addressed the extensive body of work on spectral statistics~\cite{Izrailev1988} and their connection to random matrix theory, which provides a characterization of the transition from integrable to chaotic dynamics~\cite{Haakebook} (see also \cite{Muller2026,Kieburg2026,Fyodorov2026,Novaes2026,Berkolaiko2026,Waltner2026,Lakshminarayan2026} in this volume). Although we have briefly outlined the field-theoretical approach~\cite{Altland1996,Tian2010}, particularly its connection to dynamical localization and quantum resonance through supersymmetric nonlinear $\sigma$-models, a more comprehensive treatment of this framework and its broader implications for spectral correlations and transport lies beyond the scope of this review. In addition, we have not discussed recent developments employing quantum correlation diagnostics, such as out-of-time-order correlators (see Ref.~\cite{Rozenbaum2017}, as well as Ref.~\cite{Urbina2026,garciamata2026} in this volume), which have revealed new aspects of information scrambling and operator growth in kicked systems. On the experimental side, we have also not covered recent studies that use the laser-kicked-molecules platform to realize related investigations~\cite{Bitter2017,Karle2023}. 

For variants of the kicked rotor, we have touched on only a small subset of the many generalizations explored in the literature. Modifications of the driving protocol \cite{Casati2001,Abal2002,Revuelta2024}—including alternative kicking sequences, quasi-periodic or aperiodic modulation, and more elaborate temporally structured drives—can significantly enrich the dynamical landscape. Such extensions establish connections to a range of phenomena, including discrete time crystals, anomalous transport, and forms of disorder engineered in the time domain.
In parallel, interacting extensions of the kicked rotor introduce qualitatively new physics beyond the single-particle paradigm. In particular, the kicked Bose–Einstein condensate \cite{Zhang2004,Liu2006} provides a natural and experimentally accessible bridge between quantum chaos and nonlinear mean-field dynamics. This system offers a controlled setting in which the role of interactions in shaping resonances, localization mechanisms, and dynamical stability can be systematically explored.

In the context of coupled kicked rotors, while we have discussed a concrete example at a conceptual level, a comprehensive treatment of the associated many-body phenomena lies beyond the scope of the present review and would require a separate and dedicated account. In recent years, substantial progress on many-body dynamical localization and delocalization in kicked and periodically driven systems \cite{Rylands2020,SeeToh2022,Cao2022,Guo2025} has attracted intense interest, particularly in connection with ergodicity breaking, the stability of nonthermal phases, the mechanisms of thermalization in Floquet systems, and the relation to other many-body localized states found in lattice settings
(see \cite{Abanin2019,Nandkishore2017} and \cite{zakrzewski2026} in this volume).

Looking ahead, several research directions appear particularly promising. Many-body quantum chaos in driven systems is developing rapidly (see Ref.~\cite{kolovsky2026} in this volume), and the many-body kicked rotor offers a rare opportunity to explore this regime while maintaining a clear connection to single-particle physics and semiclassical limits. Another important direction concerns dissipative quantum chaos~\cite{Carlo2005,Villasenor2024,Ferrari2025,Passarelli2025} (see also Ref.~\cite{Sa2026} in this volume). Incorporating dissipation, decoherence, or measurement into kicked systems not only brings theoretical models closer to realistic experimental conditions, but also raises fundamental questions about the fate of localization, chaos, and universality in open quantum systems. In this context, the dissipative kicked rotor may serve as a minimal model for studying the competition between coherent chaotic dynamics and environmental effects.

More broadly, the continued relevance of the kicked rotor lies in its ability to connect and reinterpret emerging concepts from diverse areas of physics. As ideas from topology, non-Hermitian physics, quantum information, and nonequilibrium statistical mechanics continue to reshape our understanding of complex quantum systems, the kicked rotor provides a flexible and conceptually transparent platform on which these ideas can be explored, tested, and unified. Far from being an exhausted model, the kicked rotor remains a vibrant paradigm in which longstanding questions acquire fresh significance and genuinely novel phenomena continue to emerge.
\appendix

\begin{ack}[Acknowledgments]

G. B. acknowledges support from INFN through the project QUANTUM. 
J. G. acknowledges support by the National Research Foundation, Singapore through the National Quantum Office, hosted in A*STAR, under its Centre for Quantum Technologies Funding Initiative (No. S24Q2d0009).
 
\end{ack}


\bibliographystyle{JHEP}%
\bibliography{refs}
\end{document}